\documentclass[11pt,letterpaper]{article}
\usepackage[T1]{fontenc}
\usepackage{lmodern,amsmath,amssymb,amsthm,mathtools,microtype}
\usepackage[margin=1in]{geometry}
\usepackage{enumitem,booktabs,tikz}
\usetikzlibrary{arrows.meta,positioning,calc}
\usepackage[hidelinks]{hyperref}
\usepackage[capitalise,nameinlink,noabbrev]{cleveref}
\crefname{algorithm}{Algorithm}{Algorithms}
\Crefname{algorithm}{Algorithm}{Algorithms}
\newtheorem{theorem}{Theorem}[section]
\newtheorem{lemma}[theorem]{Lemma}
\newtheorem{corollary}[theorem]{Corollary}

\theoremstyle{definition}

\newcommand{\OPT}{\operatorname{OPT}}
\newcommand{\dist}{\operatorname{dist}}
\newcommand{\E}{\mathbb E}
\newcommand{\Prb}{\mathbb P}
\newcommand{\Cov}{\operatorname{Cov}}
\newcommand{\supp}{\operatorname{supp}}
\newcommand{\ind}{\mathbf 1}
\newcommand{\shortsize}{s_{\mathrm{short}}}
\newcommand{\longsize}{s_{\mathrm{long}}}
\newcommand{\cycmin}{L_{\mathrm{cyc}}}
\newcommand{\hubcap}{N_{\mathrm{hub}}}
\newcommand{\hubset}{S_{\mathrm{hub}}}
\newcommand{\qv}[1]{\mathsf{#1}}
\newcommand{\joinradius}{R_{\mathrm{join}}}
\newcommand{\treeunion}{G_T}
\newcommand{\initexcess}{m_{\mathrm{excess}}}
\crefname{step}{Step}{Steps}
\Crefname{step}{Step}{Steps}
\newcounter{algorithm}
\newenvironment{procedure}[1]{\par\medskip\noindent\begin{minipage}{\linewidth}\crefalias{enumi}{step}\refstepcounter{algorithm}\hrule\vspace{5pt}\textbf{Algorithm \thealgorithm. #1}\par\smallskip\small}{\par\vspace{5pt}\hrule\end{minipage}\par\medskip}
\setlist[enumerate]{itemsep=2pt,topsep=4pt,leftmargin=*}
\setlist[itemize]{itemsep=2pt,topsep=4pt,leftmargin=*}
\hypersetup{pdftitle={Breaking the 3/2 Barrier for Minimum Strongly Connected Spanning Subgraphs},pdfauthor={MohammadTaghi Hajiaghayi and Changki Yun},pdfsubject={Deterministic approximation algorithms for directed network design}}
\title{Breaking the $3/2$ Barrier for Minimum\\Strongly Connected Spanning Subgraphs}
\author{MohammadTaghi Hajiaghayi\qquad Changki Yun\\[3pt]
  \small University of Maryland, College Park\\
  \small\texttt{\{hajiagha,ckyun\}@umd.edu}}
\date{}
\begin{document}
\maketitle

\begin{abstract}
We give a deterministic polynomial-time $(3/2-\varepsilon_0)$-approximation
for the minimum strongly connected spanning subgraph problem in unweighted
directed graphs, where $\varepsilon_0>2^{-331}$. This improves the previous ratio $3/2$ [Vetta, SODA'01; Berman, DasGupta, Karpinski, WADS'09]. The same ratio holds for the
unweighted minimum equivalent digraph problem.
\end{abstract}

\section{Introduction}\label{sec:introduction}

Given a strongly connected directed graph $G=(V,E)$, the \emph{minimum strongly
connected spanning subgraph} (SCSS) problem asks for an arc subset of minimum
cardinality that keeps $G$ strongly connected.

Let $n=|V|\ge2$ and $m=|E|$, and write $\OPT(G)$ for the minimum number of arcs
in a strongly connected spanning subgraph of $G$. Every vertex requires at
least one outgoing arc, so $\OPT(G)\ge n$. Computing $\OPT(G)$ is NP-hard,
since $\OPT(G)=n$ if and only if $G$ contains a Hamiltonian cycle~\cite{Sahni74};
in fact, SCSS remains APX-hard even when every simple directed cycle has length
at most five~\cite{BermanDK09}. On the algorithmic side, a folklore
$2$-approximation takes, for an arbitrary root $r$, a spanning out-arborescence and an in-arborescence. Their union uses at most $2n-2$
arcs and preserves strong connectivity; Frederickson and
J{\'a}J{\'a}~\cite{FredericksonJaja81} gave the analogous $2$-approximation for
weighted digraphs via minimum-cost arborescences.

Khuller, Raghavachari, and Young introduced iterative cycle contractions to
obtain a $(\pi^2/6+\varepsilon)$-approximation for every fixed
$\varepsilon>0$~\cite{KhullerRY95}, and subsequently solved the residual
instance to optimality once every cycle has length at most three, improving the
ratio to $\pi^2/6-1/36+\varepsilon$~\cite{KhullerRY96}. Vetta~\cite{Vetta01}
proposed a $3/2$-approximation based on a matching lower bound. Berman,
DasGupta, and Karpinski~\cite{BermanDK09} later pointed out gaps in Vetta's
proof and gave an alternative $3/2$-approximation based on primal--dual and
depth-first search techniques,\footnote{\Cref{app:bdk09gap} describes a gap in
one case of their charging argument; we believe that our static reachability
methods can also be adapted to close it.} while also showing that the standard
unweighted cut-LP relaxation has integrality gap at least $4/3$. Other notable
bounds include a linear-time $5/3$-approximation~\cite{ZhaoNI03} and an
LP-based $8/5$-approximation for the more general star strong connectivity
problem~\cite{Grimmer18}.

In this paper, we break the $3/2$ barrier for unweighted SCSS for the first
time.

\begin{theorem}\label{thm:main}
For a fixed constant $\varepsilon_0>2^{-331}$, there is a deterministic
polynomial-time algorithm that, given a strongly connected unweighted digraph
$G$, returns a strongly connected spanning subgraph with at most
$(3/2-\varepsilon_0)\OPT(G)$ arcs.
\end{theorem}

Any approximation ratio for unweighted SCSS extends to arbitrary digraphs via
the component decomposition of Moyles and Thompson~\cite{MoylesThompson69}. The
resulting \emph{minimum equivalent digraph} problem asks for a minimum-size arc
subset with the same transitive closure as the input digraph; for completeness,
we include a self-contained proof in \cref{subsec:equivalent}.

\begin{corollary}\label[corollary]{cor:equivalent}
There is a deterministic polynomial-time $(3/2-\varepsilon_0)$-approximation
for the unweighted minimum equivalent digraph problem.
\end{corollary}

\paragraph{Related work.}
In weighted SCSS, where each arc carries a nonnegative weight, taking a
minimum-weight in-arborescence and out-arborescence still yields a
$2$-approximation~\cite{FredericksonJaja81}, which remains the best factor
known in general. For every $\varepsilon>0$, the directed cut LP has an
instance with integrality gap at least $3/2-\varepsilon$, witnessed by the
feasible solution assigning $\frac12$ to every arc~\cite{LaekhanukitGS12}.
Hershkowitz, Kehne, and Ravi~\cite{HershkowitzKR21} round any half-integral
feasible solution to an SCSS costing at most $3/2$ times its fractional cost.
A closely related generalization is the \emph{minimum-size $k$-arc-connected
spanning subgraph} ($k$-ACSS) problem, which seeks the fewest arcs supporting
$k$ arc-disjoint directed paths between every ordered pair of vertices.
Laekhanukit, Oveis Gharan, and Singh~\cite{LaekhanukitGS12} obtained a
$\min\{7/4,\,1+1/k\}$-approximation whose guarantee depends on the number of
fractional coordinates in an extreme point of the cut LP. Using the same
directed-cut uncrossing method that bounds our arborescence prices in
\cref{lem:sparse}, we show in \cref{app:sparse-cuts} that every extreme point
of their cut LP has at most $3n-4$ fractional arcs, improving upon the $4n$
bound of Melkonian and Tardos~\cite{MelkonianTardos04} and yielding a
$\min\{5/3,\,1+3/(4k)\}$-approximation via the same rounding algorithm.

\section{Overview}\label{sec:approach}

We define \emph{matching} as set of arcs $M \subseteq E$ if each vertex has at most one incoming and outcoming edge. Our basic procedure, called \emph{matching completion}, starts from a matching with good properties and then grows a strongly connected spanning subgraph from a fixed root $r$ like an ear decomposition. It uses at most $\OPT(G)+n/2$ arcs in total, which already gives $(3/2-\Omega(1))$-approximation whenever $\OPT(G) = n + \Omega(n)$.

The entire difficulty therefore lies in the near-Hamiltonian regime $\OPT(G) = n + o(n)$. In that regime,
however, we can find such a good matching of size $n - o(n)$. A cycle containing both endpoints of many matching arcs can reduce the number of arcs we need later to connect the graph. Moreover, we can use the classic cut LP to navigate such cycles provided a large-size matching.

\subsection{\texorpdfstring{The case $\OPT(G)=n+\Omega(n)$}{The linear-excess case}}\label{subsec:rooted-matching}

Matching completion (\cref{sec:completion}) maintains a strongly connected
\emph{active set} $S\subseteq V$ containing the root $r$ and enlarges $S$ by
attaching \emph{ears}: directed paths that start and end in $S$ with all
internal vertices outside $S$. An ear through $t\ge1$ new vertices uses $t+1$ arcs, one entering each
new vertex plus one closing arc back into $S$. Absorbing one vertex at a time
can thus cost two arcs per vertex, whereas any ear through $t\ge2$ new vertices
pays at most $(t+1)/t\le 3/2$ arcs per vertex.

\begin{figure}[t]
\centering
\begin{tikzpicture}[>=Stealth,x=1cm,y=1cm,every node/.style={font=\small}]
  % Panel (a)
  \begin{scope}[shift={(0,0)}]
    \draw[rounded corners=10pt,fill=black!4,draw=black!60,thick]
      (0.1,-0.95) rectangle (2.15,1.05);
    \node[font=\small] at (0.65,0.68) {$S$};
    \node[circle,fill,inner sep=1.5pt,label=below:{$r$}] (r) at (0.75,-0.38) {};
    \node[circle,fill,inner sep=1.6pt,label=above:{$w$}] (w) at (1.65,0.42) {};
    \node[circle,fill,inner sep=1.6pt,label=above:{$u$}] (u) at (3.45,0.42) {};
    \node[circle,fill,inner sep=1.6pt,label=above:{$v$}] (v) at (5.25,0.42) {};
    \node[circle,fill,inner sep=1.4pt] (z) at (3.65,-0.55) {};
    \node[circle,fill,inner sep=1.5pt] (s2) at (1.65,-0.42) {};

    \draw[->,very thick] (w) -- node[above,font=\footnotesize] {$M$} (u);
    \draw[->,very thick] (u) -- node[above,font=\footnotesize] {$M$} (v);
    \draw[->,thick,dashed] (v) .. controls (5.15,-0.45) and (4.35,-0.55) .. (z);
    \draw[->,thick,dashed] (z) .. controls (2.95,-0.55) and (2.35,-0.45) .. (s2);
    \node[font=\footnotesize,below] at (3.75,-0.60) {return in $G-u$};
    \node[align=center,font=\footnotesize] at (3.0,-1.50)
      {\textbf{(a)} Advancing two steps along $M$ and\\closing the ear in $G-u$.};
  \end{scope}

  % Panel (b)
  \begin{scope}[shift={(7.6,0)}]
    \draw[rounded corners=10pt,fill=black!4,draw=black!60,thick]
      (0.2,-0.92) rectangle (3.1,1.02);
    \draw[rounded corners=10pt,fill=black!4,draw=black!60,thick]
      (4.4,-0.92) rectangle (7.3,1.02);
    \node[font=\footnotesize] at (1.65,-0.62) {piece $C_1\in\mathcal C$};
    \node[font=\footnotesize] at (5.85,-0.62) {piece $C_2\in\mathcal C$};

    \node[circle,fill,inner sep=1.6pt,label=above:{$a$}] (a) at (0.78,0.50) {};
    \node[circle,fill,inner sep=1.6pt,label=above:{$b$}] (b) at (2.15,0.50) {};
    \draw[->,very thick] (a) -- node[below,font=\scriptsize] {$M$} (b);

    \node[circle,fill,inner sep=1.6pt,label=below:{$u$}] (u2) at (2.58,-0.02) {};
    \node[circle,fill,inner sep=1.6pt,label=below:{$v$}] (v2) at (4.92,-0.02) {};
    \draw[->,very thick,densely dotted] (u2) -- node[above,font=\footnotesize] {$uv\in M$} (v2);

    \node[align=center,font=\footnotesize] at (3.75,-1.50)
      {\textbf{(b)} Both endpoints are covered. If $uv$ is bought,\\the pieces' attachment budgets pay for it.};
  \end{scope}
\end{tikzpicture}
\caption{Matching-guided ear attachments and budget release from pre-purchased pieces.}
\label{fig:matching-overview}
\end{figure}
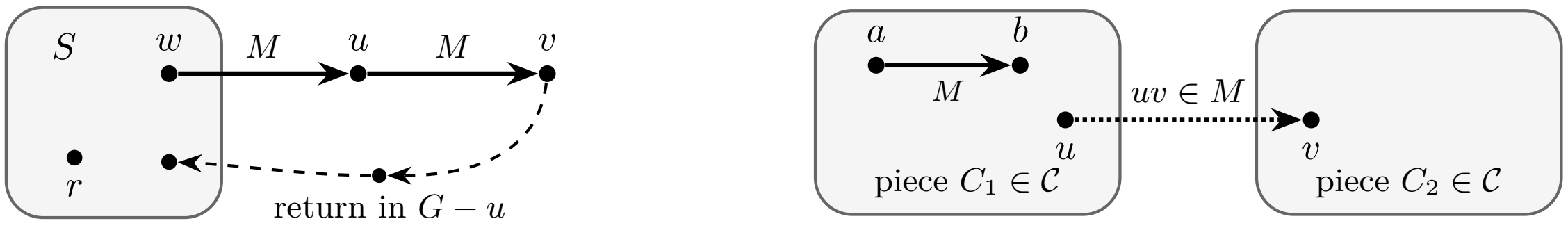

To guarantee two-step progress outside $S$, suppose an arc $wu$ leaves $S$ to a
new vertex $u$, and we wish to continue along an arc $uv$ to a second new
vertex $v$ before returning to $S$ (\cref{fig:matching-overview}(a)). Extending
$w\to u\to v$ into a simple ear requires a directed path from $v$ back to $S$
that avoids $u$; because $r\in S$, it suffices that $v$ reaches $r$ in $G-u$.
Symmetrically, growing an ear backward into $S$ along $uv$ requires that $r$
reaches $u$ in $G-v$. We call an arc $uv$ \emph{extendible at $r$} if both
conditions hold (with the forward condition automatic when $u=r$ and the
backward condition automatic when $v=r$), and write $E_r$ for the set of all
such arcs.

To chain such steps together without branching at a vertex, we want a subset
$M\subseteq E_r$ with at most one outgoing arc per tail and at most one
incoming arc per head; equivalently, $M$ is a bipartite matching between tail
copies and head copies of $V$, which we simply call a \emph{matching} in $E_r$.
Unlike a vertex-disjoint matching, a vertex $u$ may serve as both the head of
$wu\in M$ and the tail of $uv\in M$, which is precisely what lets an ear step
through $u$. In $G$, each component of $M$ is a directed path (possibly an
isolated vertex) or a directed cycle. There are exactly $n-|M|$ path components.

Every optimal solution $H^*$ of size $\OPT(G)=n+k$ contains a matching in $E_r$
of size at least $2n-\OPT(G)=n-k$. Indeed, inside $H^*$ take a spanning
in-arborescence rooted at $r$ together with one arc leaving $r$; these $n$ arcs
have distinct tails, and tree acyclicity ensures that for each arc $uv$ with
$u\ne r$, the tree path from $v$ to $r$ avoids $u$. Likewise, a spanning
out-arborescence in $H^*$ plus one arc entering $r$ gives $n$ arcs with
distinct heads passing the backward test. Their intersection inside the
$(n+k)$-arc graph $H^*$ has at least $2n-(n+k)=n-k$ arcs and is a matching in
$E_r$. Finding a maximum matching $M\subseteq E_r$ via bipartite matching
therefore leaves $n-|M|\le\OPT(G)-n=k$ path components. Berman, DasGupta, and
Karpinski~\cite{BermanDK09} also use a matching. Their completion step leaves a gap in one case of the analysis
(\cref{app:bdk09gap}). Our completion routine is incremental throughout the procedure, and stays simple in case analysis.

We use a charging scheme similar to that of Berman, DasGupta, and
Karpinski~\cite{BermanDK09}. Give each vertex a
budget of $1$ arc and each matched arc $e\in M$ an additional $\frac12$ unit.
When an ear follows two consecutive matched arcs $wu,uv\in M$ and returns in
$G-u$, every newly absorbed vertex along the ear pays $1$ for the arc entering
it, while the two $\frac12$-units on $wu$ and $uv$ pay for the single closing
arc; both matched arcs then lie inside the enlarged active set and are never
charged again. A path component of $M$ on $q$ vertices has only $q-1$ matched
arcs, leaving an unmatched start and end; giving each path $\frac12$ at each
end plus $\frac12$ in case it is first entered via a non-matching arc brings
its total allowance to $q+\frac12(q-1)+\frac32=\frac32q+1$, i.e., $\frac32$ per
vertex plus $1$ per path. A cycle component of $M$ of length $j$ has $j$ vertex
units and $j/2$ matched-arc units, paying for its $j$ arcs with $j/2$ units
left to attach to $S$. Cycles of length $j\ge4$ already have at least the $2$
units needed for an entering and a leaving arc, whereas $2$-cycles (digons) and
triangles are absorbed via a short case analysis in \cref{sec:completion} at no
extra cost. Summing over all vertices and the $n-|M|$ paths bounds the output
by
\[
  \frac32n+(n-|M|)\le \frac32n+(\OPT(G)-n)=\OPT(G)+\frac{n}{2}.
\]

\subsection{\texorpdfstring{The case $\OPT(G)=n+o(n)$}{The sublinear-excess case}}\label{subsec:near-perfect}

When $\OPT(G)=n+o(n)$, the path-end penalty $n-|M|=o(n)$ becomes negligible,
and almost all of the $3n/2$ bound comes from the $1$ unit per vertex together
with the $\frac12$ unit on each of the $n-o(n)$ matched arcs. Suppose that,
before running matching completion, we buy $m'$ original arcs forming a family
$\mathcal C$ of vertex-disjoint strongly connected subgraphs (\emph{pieces})
with vertex union $V_{\mathcal C}$, and allocate $2$ units per piece to attach
it to $S$.

\paragraph{Piece excess and covered matching arcs.}
The vertex budgets pay for all arcs of a directed cycle $C$. Deleting its
vertices from $M$ can be costly. If they alternate with $|V(C)|$ outside vertices
on a matching cycle, deletion cuts $2|V(C)|$ matched arcs and leaves $|V(C)|$
isolated vertices. Matching completion keeps $M$ on the original graph.
Once both endpoints of $uv\in M$ lie in $V_{\mathcal C}$, its half unit is
free. The endpoints can even belong to different pieces. If completion uses
$uv$ to connect those pieces, their attachment budgets pay for the arc
(\cref{fig:matching-overview}(b)). For a single purchased cycle containing
both endpoints of $q$ matched arcs, we reserve two units for attachment and free
$q/2$ relative to completion with the same $M$ and no pieces.

More generally, write $E[W]$ for the input arcs with both endpoints in $W$ and
$y(F)=\sum_{e\in F}y_e$ for the weight of an arc set $F$. A \emph{fractional
matching} $y\in\mathbb R_{\ge0}^{E_r}$ assigns nonnegative weights to $E_r$
summing to at most $1$ at each tail and at each head. Once $\mathcal C$ is
fixed, matching completion selects an integral matching $M\subseteq E_r$
maximizing $2|M|+|M\cap E[V_{\mathcal C}]|$. Because the bipartite matching
polytope is integral, $M$ does at least as well as any fractional matching $y$
(\cref{cor:fractional}), yielding an output $E_{\rm out}$ of size
\[
 |E_{\rm out}|\le \frac32 n+
 \underbrace{\bigl(m'-|V_{\mathcal C}|+2|\mathcal C|\bigr)}_{\text{piece excess}}
 +\underbrace{\bigl(n-y(E)\bigr)}_{\text{matching deficit}}
 -\underbrace{\tfrac12 y(E[V_{\mathcal C}])}_{\text{budget freed}} .
\]
Buying cycles is worthwhile when the matching budget they free exceeds their
piece excess. If it also pays the matching deficit with $\Omega(n)$ to spare,
completion beats $3n/2$ by a linear amount.

\paragraph{Covering matching weight via long-cycle contraction.}
How can we find cycles that enclose substantial matching weight? A single
extreme-point optimum $x$ of the directed cut LP,
\begin{equation}\label{eq:cut-lp}
 \min\bigl\{x(E)\mid x\in\mathbb R_{\ge0}^E,\quad x(\delta^-(A))\ge1
                 \quad(\varnothing\ne A\subsetneq V)\bigr\},
\end{equation}
where $\delta^-(A)$ is the set of arcs entering $A$, both supplies a
near-perfect fractional matching and guides our cycle search. Clearly
$n\le x(E)\le\OPT(G)=n+o(n)$. Applying the cut constraints to the sets of
vertices reaching $r$ in $G-u$ (and symmetrically in the reverse graph)
extracts a fractional matching $x'\le x$ on $E_r$ of weight
$x'(E)\ge 2n-x(E)=n-o(n)$ (\cref{lem:cut-matching}). Taking $y=x'$ gives
matching deficit $n-x'(E)=o(n)$, and since $(x-x')(E)=o(n)$, the $x$-weight of
any arc set tracks its $x'$-weight up to $o(n)$.

By directed cut uncrossing (\cref{lem:sparse}), the extreme point $x$ has at
most $3n-4$ positive coordinates. Give each arc $e$ in the support $\supp x$
price $p_e=1/x_e$. Because the fractional cost is
$\sum_{e\in\supp x}p_e x_e=|\supp x|<3n$, Edmonds' branching packing theorem
yields a spanning in-arborescence $T^{\rm in}$ and a spanning out-arborescence
$T^{\rm out}$ rooted at $r$, each of price less than $3n$. In their union
$\treeunion=T^{\rm in}\cup T^{\rm out}$, the reciprocals sum to at most $6n$,
so Cauchy--Schwarz forces every subset $A\subseteq E(\treeunion)$ to
carry quadratic $x$-weight:
\[
 |A|^2 \le x(A)\sum_{e\in A}\frac{1}{x_e} \le 6n\,x(A)
 \quad\Longrightarrow\quad
 x(A)\ge \frac{|A|^2}{6n}.
\]

\emph{Long-cycle contraction} (\cref{sec:long}) maintains a quotient digraph
$\widehat G$ obtained by contracting the disjoint pieces bought so far into
single vertices (discarding loops and keeping parallel crossing arcs) and
tracks the strongly connected spanning image $F=\pi(\treeunion)$ under the
projection $\pi:V(G)\to V(\widehat G)$. At each step it builds a
polynomial-size list of simple cycles in $\widehat G$, each using at most one
arc outside $F$. We fix a length threshold $\cycmin=O(1)$. Whenever the
list contains a cycle of length $j\ge\cycmin$, we buy representative input
arcs, choosing at least $j-1$ arcs from $\treeunion$, and contract the cycle.
Each contraction reduces $|V(\widehat G)|$ by $j-1\ge\cycmin-1$, absorbs at
least $j-1$ fresh arcs of $\treeunion$ (which become loops and are never bought
again), and adds at most $2$ to the piece excess. Writing
$n_{\rm drop}=n-|V(\widehat G)|$ for the cumulative drop in vertex count, the
linear piece excess competes against the quadratic freed budget:
\[
 |E_{\rm out}|\le \frac32 n
       +\frac{2n_{\rm drop}}{\cycmin-1}
       -\frac{n_{\rm drop}^2}{12n}
       +o(n).
\]
Once $n_{\rm drop}$ reaches the contraction threshold of $\Theta(n/\cycmin)$,
the quadratic gain dominates and matching completion finishes with at most
$\frac32n-\Omega(n/\cycmin^2)$ arcs. Otherwise the list runs out of cycles of
length at least $\cycmin$ early, leaving a residual quotient $\widehat G$ whose
purchased pieces cover only $O(n/\cycmin)$ original vertices at piece excess
$O(n/\cycmin^2)$.

\paragraph{Local structure, hubs, and exact solutions for far pieces.}
Suppose long-cycle contraction halts early. Our cycle list in $\widehat G$
(\cref{subsec:cycle-list}) pairs every quotient arc $e$ with a return path
$P_e\subseteq F$ to form $C_e=e+P_e$. For $e\in E(F)$ it also tests cycles
obtained by combining $C_e$ with paths in a depth-first search tree
$T_{\rm DFS}$ of $F$. Because every listed cycle now has length less
than $\cycmin$, \cref{sec:decomposition} constructs a tree decomposition of the
underlying undirected graph of $\widehat G$, indexed by the nodes of
$T_{\rm DFS}$, in which the bag at each node $\qv v$ lies inside a strongly connected
\emph{enclosure} $U_{\qv v}$ of round-trip radius at most $5\cycmin$; here the
round-trip distance $d_{\rm rt}(\qv u,\qv v)$ is the shortest length of a
closed directed walk through $\qv u$ and $\qv v$ in $\widehat G$.

Bounded round-trip radius in $\widehat G$ does not automatically bound the
number of original vertices in $\pi^{-1}(U_{\qv v})$. However, if we traverse a
closed walk of length $O(\cycmin)$ in reverse and replace each reversed arc $e$
by its forward return path $P_e\subseteq F$ of length less than $\cycmin$, we
obtain a directed walk entirely inside $F$ of length $O(\cycmin^2)$. Such walks
can reach many original vertices only by visiting a quotient vertex $\qv v$
that either has high degree in $F$ (unbounded branching) or represents a large
contracted piece $\pi^{-1}(\qv v)$. Defining the set of \emph{hubs} $\hubset$
to consist of quotient vertices with $\deg_F(\qv v)>\hubcap$ or
$|\pi^{-1}(\qv v)|>\hubcap$, we obtain both $|\hubset|\le 5n/\hubcap$ (since
$|E(F)|\le 2n$ and preimages partition $V$) and a constant bound on
$|\pi^{-1}(U_{\qv v})|$ whenever $\qv v$ has round-trip distance greater than
$6\cycmin$ to every hub.

For a far-vertex fraction $\xi$, the separator argument uses a round-trip
radius $R=\Theta(\cycmin/\xi)$ (\cref{eq:radius}). If at least $\xi n$
original vertices project beyond this radius from $\hubset$, \emph{separator contraction}
(\cref{subsec:local-solutions}) isolates bounded-size far subgraphs and solves
them optimally. By averaging over concentric distance bands around $\hubset$
and residue classes of $T_{\rm DFS}$-depth on the far side of the band (which
slice deep bounded-degree subtrees into $O(1)$-size components), we mark a set
of bags whose merged strongly connected enclosures $Z\subseteq V(G)$ contain at
most $\xi n/16$ original vertices. Contracting these enclosures in $G$ reduces
every marked separator bag to one vertex in the resulting graph $G'$,
so adjacent pieces share at most one vertex and
$\OPT(G')=\sum_D\OPT(D)\le\OPT(G)$. We solve the bounded-size far pieces
$\mathcal D_{\rm ex}$ to optimality by enumeration, run rooted matching
completion ($\le\frac32\OPT(D)$) on all other pieces, and lift the union back
to $G$ by adding an in- and out-arborescence inside each contracted component
of $Z$ at cost at most $2|Z|$:
\[
 |E_{\rm out}|
 \le \sum_{D\in\mathcal D_{\rm ex}}\OPT(D)
     +\frac32\sum_{D\notin\mathcal D_{\rm ex}}\OPT(D)+2|Z|
 \le \frac32\OPT(G)-\frac12\sum_{D\in\mathcal D_{\rm ex}}\OPT(D)+2|Z|.
\]
Every far vertex outside $Z$ has an outgoing arc inside its unique piece in
$\mathcal D_{\rm ex}$, so $\sum_{D\in\mathcal D_{\rm ex}}\OPT(D)\ge\xi n-|Z|$,
yielding a net saving of at least $\frac14\xi n$ arcs below $\frac32\OPT(G)$.

\paragraph{Shared returns via the hub forest and fractional path covers.}
In the remaining case, at each block size $s\in\{\shortsize,\longsize\}$ fewer
than $\xi_s n$ original vertices project farther than $R_s$ from $\hubset$.
We buy a \emph{hub forest} connecting nearby hubs in both directions.
It has at most $|\hubset|\le 5n/\hubcap$ links, each of round-trip length at
most $\joinradius=O(1)$. With the hub threshold of \cref{sec:parameters},
the resulting pieces have vertex union $U$ of size $O(n/\cycmin)$ and piece
excess $\initexcess=O(n/\cycmin^2)$ (\cref{subsec:shared-returns}).

Now consider any $s$-vertex directed path $P$ from $u$ to $v$ in $G$ whose
endpoints project within round-trip distance $R_s$ of hubs $\qv h_u,\qv h_v$.
The projection of $P$ has at most $s-1$ arcs, each with a return of length
less than $\cycmin$. The triangle inequality puts $\qv h_u$ and $\qv h_v$ in
the same hub-forest component. We close $P$ by routing
$\pi(v)\leadsto\qv h_v$, traversing the already-bought
hub-forest piece to $\qv h_u$ without buying new arcs, and routing
$\qv h_u\leadsto\pi(u)$. The two endpoint routes
cost at most $2R_s$ new arcs. If all $s$ vertices of $P$ are new, their vertex
budgets pay for the $s-1$ path arcs and one return arc, so the piece excess
increases by at most $2R_s-1$. When $P$ revisits earlier pieces, only its fresh
vertices bring new budget.

To supply such paths across $G$, observe that omitting the incoming root arc
from the tree intersection in \cref{subsec:rooted-matching} leaves an acyclic
matching in $E_r$ partitioning $V$ into at most $k+1=o(n)$ directed paths;
chopping them at a random offset into $\longsize$-vertex blocks loses few
vertices to incomplete ends or far endpoints. Without knowing an optimum, the
explicit LP of \cref{subsec:path-covers} computes a \emph{fractional path
cover} directly in $G$, assigning weights to $\longsize$-vertex simple paths in
$E_r$ with hub-near endpoints and to uncovered \emph{singletons} so that each
vertex receives total covering weight $1$ while minimizing total singleton
weight $b_{\mathrm{long}}^{(1)}$.

The cut matching $x'$ need not align with these paths. The long cover supplies
its own fractional matching $y$ on $E_r$: let $y_e$ be the total weight of
paths using $e$. Every block has one fewer arc than vertices, so the matching
deficit equals the total block weight $b_{\mathrm{long}}$. The singleton
weight $b_{\mathrm{long}}^{(1)}$ accounts for one vertex per unit, and the
other blocks cover $\longsize$ vertices per unit. Thus
\[
 n-y(E)=b_{\mathrm{long}}=\frac{n}{\longsize}
          +\left(1-\frac1{\longsize}\right)b_{\mathrm{long}}^{(1)}.
\]
Even a cover with no singletons leaves a linear deficit $n/\longsize$. The
freed matching budget must pay for this deficit and the piece excess.

\paragraph{Overlap cancellation and the two-scale trade-off.}
Given a fractional cover of block size $s$ and total block weight $b_s$,
\emph{shared-return rounding} (\cref{subsec:block-construction,sec:final-analysis})
draws $N_s=\lfloor b_s/2\rfloor$ blocks independently in proportion to their
weights, closes the drawn paths through the hub forest, and completes the
resulting pieces. Each vertex has covering weight one, so its probability of
appearing in at least one draw is $\beta=1-(1-1/b_s)^{N_s}$.

Let $S_i$ be $U$ together with the vertices of the first $i$ drawn blocks.
If the next path leaves $S_i$ and re-enters it after $j$ fresh vertices, the
run uses $j+1$ arcs. We charge at most one extra arc for the run, half to each
boundary crossing. At either block size, the
paths use an arc $e$ to total weight at most $y_e$. Hence the expected charge
for these runs is at most $y(\delta(S_i))/(2b_s)$. Each arc crossing $S_i$
has one endpoint still outside, which the next draw hits with probability
$1/b_s$. The matching budget that this frees pays the expected charge.

A draw can also cover both endpoints of an arc that was wholly outside $S_i$.
For each $s$, let $Z_s$ sum, over arcs $uv$ outside $U$, the matching weight
$y_{uv}$ times the total weight of blocks in the size-$s$ cover containing
both $u$ and $v$. These joint hits give the long cover a further expected gain
of at least $Z_{\mathrm{long}}/12$ in the completion bound. In particular,
every long path that uses $uv$ contains both endpoints, so
$Z_{\mathrm{long}}\ge\|w\|_2^2$, where $w$ is the restriction of $y$ to arcs
with both endpoints outside $U$. Subdividing the long paths at all offsets
gives the short cover and preserves $Z_{\mathrm{short}}\le Z_{\mathrm{long}}$.

The short cover exploits matching weight spread over many arcs. Write $X$
for the vertices hit by its draws,
so $\E|X|=\beta n$. If endpoints were sampled independently, an arc would
cross $\delta(X)$ with probability $2\beta(1-\beta)$ and lie inside $X$ with
probability $\beta^2$. For the actual cover, \cref{subsec:block-gains} bounds
the correlations using the fact that one short block contains at most
$\shortsize$ vertices. We discard the crossing arcs of $w$, rescale its
weights inside $X$ by $1/\beta$ and inside the complement by $1/(1-\beta)$,
and trim rows and columns to capacity one. The resulting matching $y'$ has
$y'(E[X])\ge |X|-(n-y'(E))$. The budget freed inside $X$ is therefore at
least half the sampled vertex count minus half the matching deficit. The sum of the variances of the
sampled weights at heads and tails is at most $\shortsize\|w\|_2^2$.
By Cauchy--Schwarz the expected trimming loss is
$O(\sqrt{n\shortsize\|w\|_2^2})$. Since $\|w\|_2^2\le Z_{\mathrm{long}}$,
balancing this bound against the short cover's linear gain gives a threshold
of $\Theta(n/\shortsize)$ for $Z_{\mathrm{long}}$. Below the threshold the
short cover retains $\Omega(n)$ after correlation losses. At or above it,
joint hits give the long cover $\Omega(n/\shortsize)$.

For short blocks, $\xi_{\mathrm{short}}=\Theta(1)$ gives
$R_{\mathrm{short}}=\Theta(\cycmin)$. We choose
$\shortsize=\Theta(R_{\mathrm{short}})$ so that the short cover still gains
$\Omega(n)$ after paying for returns and singletons below the threshold.
For long blocks, $\xi_{\mathrm{long}}=\Theta(1/\cycmin^2)$ bounds the
singleton weight from far endpoints by $O(n/\cycmin^2)$, the order of the
initial piece excess. This gives $R_{\mathrm{long}}=\Theta(\cycmin^3)$,
so $\longsize=\cycmin^5$ puts the return cost
$R_{\mathrm{long}}n/\longsize$ on the same scale. At this fixed block length
the $o(n)$ paths in the optimal solution's path partition lose only $o(n)$
vertices to incomplete ends. Thus the long cover pays $O(n/\cycmin^2)$
against a gain of $\Omega(n/\cycmin)$ at the threshold
(\cref{sec:parameters}).

We choose the block size with the smaller expected upper bound, fix its
draws by conditional expectations (\cref{subsec:derandomize}), and compare
the resulting subgraph with the rooted solution.

\section{Preliminaries}\label{sec:preliminaries}

For unweighted SCSS and minimum equivalent digraphs we discard self-loops and
keep one input arc per ordered pair. For $k$-ACSS we retain parallel input arcs,
since they can serve different arc-disjoint paths. By convention, every
vertex reaches itself by a path of length zero, so $\OPT(H)=0$ when
$|V(H)|=1$. Write $uv$ for the directed arc from tail $u$ to head $v$. Unless
specified otherwise, walks, paths, and cycles are directed; paths and cycles
are simple, their length counts arcs, and a \emph{closed walk} starts and ends
at the same vertex. For $S\subseteq V$, write $G[S]$ for the subgraph induced
by $S$, $E[S]$ for the set of arcs with both endpoints in $S$, and
$G-u=G[V\setminus\{u\}]$.

When we contract a collection of vertex-disjoint strongly connected subgraphs
of $G$ into single vertices to form a \emph{quotient} digraph $\widehat G$ with
projection $\pi:V(G)\to V(\widehat G)$, we discard the resulting self-loops and
retain each crossing arc together with its identity in $G$ (including its
original tail and head in $V(G)$). Distinct input arcs may therefore become
parallel arcs in $\widehat G$. For any subgraph $H$ of $G$, write $\pi(H)$ for
the subgraph of $\widehat G$ on $\pi(V(H))$ formed by the non-loop images of
$E(H)$. When a graph and its quotient appear together, we write vertices of the
quotient in sans-serif font (e.g., $\qv u,\qv v$).

\subsection{Directed cuts, arborescences, and matchings}

For $S\subseteq V$, write $\delta^-(S)$ and $\delta^+(S)$ for the sets of arcs
entering and leaving $S$, respectively, and let
$\delta(S)=\delta^-(S)\cup\delta^+(S)$ denote the set of arcs with exactly one
endpoint in $S$. For a vector $x\in\mathbb R^E$ and $F\subseteq E$, write
$x(F)=\sum_{e\in F}x_e$ and $\supp x=\{e\in E:x_e>0\}$, and write
$\ind_{\{\mathcal E\}}$ for the $\{0,1\}$-indicator of an event or condition
$\mathcal E$. For an integer $h\ge1$, define the directed cut polyhedron
\[
 P_h=\{x\in\mathbb R_{\ge0}^{E}:
        x(\delta^-(S))\ge h\quad(\varnothing\ne S\subsetneq V)\}.
\]
The SCSS cut relaxation~\labelcref{eq:cut-lp} minimizes $x(E)$ over $P_1$.
Evaluating singleton cuts $\delta^-(\{v\})$ on the one hand and the indicator
of an optimal SCSS on the other gives $n\le \min_{x\in P_1}x(E)\le\OPT(G)$.

To separate over $P_h$, fix a root $r$: after checking $x\ge0$, treat $x_e$ as
arc capacities and compute minimum $s$--$t$ cuts in both orientations between
$r$ and every $v\ne r$. Because every nonempty proper subset $S\subsetneq V$
either excludes $r$ (separating $r$ from each $v\in S$) or contains $r$
(separating each $v\notin S$ from $r$), $x\in P_h$ holds if and only if all
$2(n-1)$ minimum cuts have capacity at least $h$. Since $P_h$ lies in the
nonnegative orthant (hence contains no lines) and the optimal face
of~\labelcref{eq:cut-lp} is bounded, the ellipsoid method with this separation
oracle computes an extreme-point optimum in polynomial time~\cite{GLS88} (and
similarly over $P_k\cap[0,1]^E$). We also solve the explicit polynomial-size
linear program~\labelcref{eq:path-lp} in polynomial time.

\begin{lemma}[Sparse directed cuts]\label[lemma]{lem:sparse}
Every extreme point of $P_h$ on $n\ge2$ vertices has at most $3n-4$
positive coordinates. Every extreme point of $P_h\cap[0,1]^E$ has at
most $3n-4$ coordinates strictly between zero and one.
\end{lemma}

\Cref{app:sparse-cuts} proves \cref{lem:sparse} and applies its second part to
the $k$-ACSS problem discussed in \cref{sec:introduction}.

\paragraph{Arborescences.}
Recall that an out-arborescence rooted at $r$ is a spanning tree directed away
from $r$, and an in-arborescence is directed toward $r$. Given rational
$x\in P_1$ and nonnegative arc prices $p_e$ defined on $\supp x$, there exist
an out-arborescence and an in-arborescence rooted at $r$, both supported on
$\supp x$, each of total price at most $\sum_{e\in\supp x} p_e x_e$. Indeed,
clearing denominators of $x$ by a common integer $d\ge1$ and replacing each arc
$e\in\supp x$ by $dx_e$ parallel copies produces a multigraph in which every
nonempty $S\subseteq V\setminus\{r\}$ has at least $d$ entering arcs. Edmonds'
branching packing theorem~\cite{packing} yields $d$ arc-disjoint
out-arborescences rooted at $r$ of combined price at most
$d\sum_{e\in\supp x}p_e x_e$, so the cheapest has price at most
$\sum_{e\in\supp x}p_e x_e$; applying the same argument in the reverse digraph
(using $\delta^-(V\setminus S)=\delta^+(S)$) bounds the minimum-price
in-arborescence. Minimum-price arborescences on $\supp x$ can be computed in
polynomial time~\cite{Edmonds67,GabowGST86}.

\paragraph{Matchings.}
Identifying each arc $uv\in E$ with an undirected edge between a tail copy of
$u$ and a head copy of $v$ yields a bipartite graph on $V\sqcup V$. A
\emph{fractional matching} is a vector $y\in\mathbb R_{\ge0}^E$ satisfying
$y(\delta^+(u))\le 1$ and $y(\delta^-(u))\le 1$ for every $u\in V$; an integral
\emph{matching} $M\subseteq E$ is an arc subset in which no two arcs share a
tail and no two share a head. The extreme points of the fractional matching
polytope are integral, and a maximum-weight matching can be found in polynomial
time~\cite{Schrijver03}.

\subsection{Minimum equivalent digraphs}\label{subsec:equivalent}

An arc subset $H\subseteq E$ is \emph{equivalent} to $G$ if $H$ and $G$ have
the same reachability relation on $V\times V$; write $\OPT_{\rm eq}(G)$ for the
minimum cardinality of such an $H$. The component reduction of Moyles and
Thompson~\cite{MoylesThompson69} transfers any SCSS approximation guarantee to
minimum equivalent digraphs.

\begin{proof}[Proof of \cref{cor:equivalent}]
Compute the family $\mathcal C$ of strongly connected components of
$G$~\cite{Tarjan72}. Contracting each component in $\mathcal C$ produces an
acyclic \emph{condensation} digraph, inducing a strict reachability partial
order $C<C'$ ($C\ne C'$ and $G$ contains a directed path from $C$ to $C'$). A
comparable pair $C<C'$ is a \emph{cover relation} if no component $C''$
satisfies $C<C''<C'$. Every directed path from $C$ to $C'$ realizing a cover
relation must consist of a single intercomponent arc from $C$ to $C'$ together
with paths inside $C$ and $C'$; conversely, retaining one direct arc for each
cover relation preserves all reachability between distinct components.

Because no directed path can leave a strongly connected component and later
return to it, any equivalent subgraph $H$ must induce a strongly connected
spanning subgraph on every $C\in\mathcal C$. Conversely, the union of a
strongly connected spanning subgraph on each $C\in\mathcal C$ with one
representative arc for each of the $c$ cover relations is equivalent to $G$,
giving the exact identity
\[
 \OPT_{\rm eq}(G)=c+\sum_{C\in\mathcal C}\OPT(G[C]),
\]
where $\OPT(G[C])=0$ when $|C|=1$. Compute the cover relations via transitive
closure, pick one representative arc for each, and apply \cref{thm:main} to
every component $C\in\mathcal C$ with $|C|\ge2$. This buys the $c$
intercomponent arcs optimally and approximates each component term within
$3/2-\varepsilon_0$.
\end{proof}

\section{Matching completion with prescribed pieces}\label{sec:completion}

Starting from a fixed root $r$, we first show how a matching guides ear
attachments to build a strongly connected spanning subgraph of $G$ with at most
$\OPT(G)+n/2$ arcs. When $\OPT(G)=n+k$, this saves at least $k/2$ arcs relative
to $3\OPT(G)/2$, giving a linear improvement whenever $k=\Omega(n)$. We keep
this rooted solution for comparison with the candidates produced in
\cref{sec:long,sec:decomposition,sec:near-hubs}. After contracting long cycles
or closing paths through hubs, we encounter the more general problem in which a
disjoint family of strongly connected pieces has already been purchased, and
all arcs bought inside those pieces must be retained in the final subgraph.

An arc $uv\in E$ is \emph{extendible at $r$} if it satisfies both reachability
conditions
\begin{equation}\label{eq:root-tests}
 \begin{aligned}
 u=r &\quad\text{or $v$ reaches $r$ in $G-u$},\\
 v=r &\quad\text{or $r$ reaches $u$ in $G-v$}.
 \end{aligned}
\end{equation}
Write $E_r\subseteq E$ for the set of arcs extendible at $r$. The paths
witnessing~\labelcref{eq:root-tests} may use arbitrary arcs of $G$ outside the
deleted vertex.

Given a matching $M\subseteq E_r$, every vertex in the spanning subgraph
$(V,M)$ has indegree at most one and outdegree at most one, so the weakly
connected components of $(V,M)$ are vertex-disjoint directed cycles and
directed paths (including isolated vertices as length-zero paths). Each path
component has one more vertex than arc, and therefore $(V,M)$ has exactly
$n-|M|$ path components; each path has one \emph{missing predecessor} at its
start vertex and one \emph{missing successor} at its end vertex (both at the
same vertex if the path is an isolated vertex).

Any optimal solution $H^*$ of size $\OPT(G)$ contains a matching in $E_r$ of
size at least $2n-\OPT(G)$. Indeed, inside $H^*$ take a spanning
in-arborescence rooted at $r$ together with one arc of $H^*$ leaving $r$: these
$n$ arcs have distinct tails, and for each arc $uv$ with $u\ne r$, acyclicity
of the in-arborescence ensures that its unique path from $v$ to $r$ avoids $u$,
verifying the first condition of~\labelcref{eq:root-tests}. Symmetrically, a
spanning out-arborescence in $H^*$ together with one arc entering $r$ gives $n$
arcs with distinct heads satisfying the second condition. Their intersection in
$H^*$ has at least $2n-\OPT(G)$ arcs and forms a matching in $E_r$, so a
maximum matching $M\subseteq E_r$ leaves $n-|M|\le\OPT(G)-n$ path components.

\paragraph{Matching completion.}\label{proc:matching-completion}
Given a family $\mathcal C$ of vertex-disjoint selected strongly connected
subgraphs of $G$ with vertex union $V_{\mathcal C}$, compute a maximum-weight
matching $M\subseteq E_r$ by giving each arc in $E_r\cap E[V_{\mathcal C}]$
weight $3$ and every other arc in $E_r$ weight $2$ (when
$\mathcal C=\varnothing$, this simply chooses a maximum-cardinality matching in
$E_r$). Merge each directed cycle of $M$ that intersects $V_{\mathcal C}$ with
all members of $\mathcal C$ it meets, and then repeatedly attach ears to the
part containing $r$ according to the rules in the proof of
\cref{lem:completion} until a single strongly connected set spans $V$. Return
all selected original arcs, retaining every arc previously bought inside
$\mathcal C$. As shown below, the bound~\labelcref{eq:completion} holds for
every matching $M\subseteq E_r$ supplied to these attachment steps.

\begin{lemma}[Completion estimates]\label[lemma]{lem:completion}
There is a deterministic polynomial-time algorithm returning a
strongly connected spanning subgraph of $G$ with at most $\OPT(G)+n/2$ arcs.
More generally, let $\mathcal C$ be a family of vertex-disjoint selected
strongly connected subgraphs of $G$, let $V_{\mathcal C}$ be the union of their
vertex sets, and let $m'$ be the total number of original arcs selected in
$\mathcal C$. Given any matching $M\subseteq E_r$, a deterministic
polynomial-time algorithm returns a strongly connected spanning subgraph of $G$
containing every selected arc of $\mathcal C$ and having at most
\begin{equation}\label{eq:completion}
 \frac32 n+\bigl(m'-|V_{\mathcal C}|+2|\mathcal C|\bigr)
 +(n-|M|)-\frac12|M\cap E[V_{\mathcal C}]|
\end{equation}
arcs. We call $m'-|V_{\mathcal C}|+2|\mathcal C|$ the \emph{piece excess} of
$\mathcal C$.
\end{lemma}

\begin{proof}
We grow a strongly connected \emph{active set} initialized to the part
containing $r$. An \emph{ear} relative to the active set is either a directed
path with both endpoints in the active set and all internal vertices outside
it, or a simple directed cycle meeting the active set in a single vertex;
adding its arcs keeps the active set strongly connected.

Throughout the process we maintain a partition of $V$ into strongly connected
parts and work in the corresponding quotient (retaining original endpoints and
parallel crossing arcs while discarding self-loops, as in
\cref{sec:preliminaries}). An \emph{attachment} is a simple directed cycle
through the active part in this quotient. Because each visited part is already
strongly connected via its selected internal arcs, adding the quotient cycle's
connecting input arcs (together with the internal arcs of any newly bought
matching cycle) merges all visited parts into the active set while retaining
all previously selected arcs~\cite{KhullerRY95}. If an attachment absorbs $t$
new quotient parts, it uses $t+1$ connecting arcs: one \emph{ordinary
connector} for each new part, plus one \emph{extra connector} to close the
cycle back into the active part.

Beyond the $m'$ arcs already bought inside $\mathcal C$, we distribute an arc
budget across the instance as follows:
\begin{itemize}
\item $2$ units to each piece in $\mathcal C$ (total $2|\mathcal C|$);
\item $1$ unit to each vertex outside $V_{\mathcal C}$ (total
  $n-|V_{\mathcal C}|$);
\item $\frac12$ unit to each matched arc in $M\setminus E[V_{\mathcal C}]$,
  i.e., with at least one endpoint outside $V_{\mathcal C}$ (total
  $\frac12(|M|-|M\cap E[V_{\mathcal C}]|)$); and
\item $\frac32$ units to each of the $n-|M|$ path components of $(V,M)$:
  $\frac12$ for its missing predecessor end, $\frac12$ for its missing
  successor end, and $\frac12$ reserved for its first attachment (total
  $\frac32(n-|M|)$).
\end{itemize}

\paragraph{Merging matching cycles that meet $\mathcal C$.}
Before growing from $r$, we merge every cycle of $M$ that intersects
$V_{\mathcal C}$ with all members of $\mathcal C$ it meets, leaving every
remaining cycle of $M$ disjoint from $V_{\mathcal C}$ (\emph{intact}). Take the
union of the subgraphs in $\mathcal C$ and the cycles of $M$ meeting
$V_{\mathcal C}$. Each connected component of this union is a connected union
of strongly connected subgraphs and is therefore strongly connected. Inside
each such component, retain all arcs already selected in $\mathcal C$, contract
each member of $\mathcal C$ to a single vertex, and choose a spanning
in-arborescence and a spanning out-arborescence rooted at a common vertex in
the resulting component quotient.

Suppose this component merges $c\ge1$ pieces of $\mathcal C$. Each of the two
arborescences induces an acyclic forest on the $c$ contracted piece vertices
and hence uses at most $c-1$ arcs between them; buying those arcs costs at most
$2(c-1)$ units out of the $2c$ piece units, leaving $2$ units for the newly
merged piece. Every other arc belonging to the union of the two arborescences
is bought once and is a cycle arc of $M$ with at least one endpoint outside
$V_{\mathcal C}$: it carries its own $\frac12$-unit budget, and because each
cycle vertex outside $V_{\mathcal C}$ is incident to at most two cycle arcs,
that vertex's $1$-unit budget supplies the remaining $\frac12$-unit for each
incident cycle arc.

After these mergers, $V$ is partitioned into (i) the resulting \emph{selected
pieces} (which may include $1$-vertex members of $\mathcal C$), (ii)
\emph{intact matching cycles} disjoint from $V_{\mathcal C}$, and (iii)
\emph{ordinary singletons}, namely the vertices outside $V_{\mathcal C}$ that
lie on path components of $(V,M)$. We refer to selected pieces and intact
matching cycles collectively as \emph{whole parts}. Note that an intact cycle
of length $j\ge2$ holds $j$ vertex units plus $j/2$ matched-arc units ($3j/2$
in total), enough to buy all $j$ of its arcs together with its $1$ ordinary
connector, with $(j-2)/2$ units to spare. Initialize the active set to the part
containing $r$ (buying its $j$ arcs if it is an intact cycle; its remaining
budget is unused since $r$ needs no connectors).

At each step while the active set is proper, we apply the first applicable rule
below:
\begin{enumerate}
\item If a matched path arc crosses the boundary of the active set, follow it
  using the matching-arc rule below.
\item Otherwise every path of $M$ lies wholly inside or wholly outside the
  active set. If an input arc joins the active set (in either direction) to an
  ordinary singleton $u$, use that arc in place of $wu$ in the matching-arc
  rule below, paying the missing $\frac12$-unit from that path's
  first-attachment $\frac12$-unit; once $u$ is absorbed, the path now meets the
  active set, so any remaining vertices of that path will subsequently be
  reached via Rule~1 and its first-attachment $\frac12$-unit is never needed
  again.
\item Otherwise every part adjacent to the active set in the quotient is a
  whole part; apply the whole-part rules below.
\end{enumerate}

\paragraph{Attachments following a matching arc.}
Suppose a matched arc $wu\in M$ leaves the active set and enters an ordinary
singleton $u$. If $u$ has a matched successor $uv\in M$, begin with $wu$ and
$uv$. If $v$ already lies in the active set, $wu$ and $uv$ alone close an
attachment. Otherwise, extendibility of $uv$ gives a directed path from $v$ to
$r$ in $G-u$: stop this path at its first vertex in the active set, project the
prefix onto the current quotient (treating each whole part as one vertex), and
shortcut any repeated quotient vertices along the walk to obtain a simple
quotient path back to the active part avoiding $u$. Retain $wu$, $uv$, and one
input arc for each quotient arc along the return, buying the internal arcs of
any intact matching cycle visited. Every newly absorbed ordinary singleton pays
$1$ for its entering arc, every visited whole part pays its ordinary cost ($1$
ordinary connector, plus $j$ internal arcs for an intact cycle), and the two
$\frac12$-units on $wu$ and $uv$ pay for the extra connector. If instead $u$
has no matched successor, return to the active set from $u$ along any simple
quotient path; the $\frac12$-unit on $wu$ together with the path's
missing-successor $\frac12$-unit pays for the extra connector. By symmetry, the
second condition of~\labelcref{eq:root-tests} gives the reverse construction
whenever a matched arc from an ordinary singleton enters the active set.

In every case above, all matched arcs and path-end allowances spent in the step
become strictly internal to the enlarged active set and are never charged
again. Finally, if the crossing matched arc $wu$ leads directly between the
active set and a selected piece (rather than an ordinary singleton), we simply
close a shortest quotient cycle through that piece: the piece's $2$-unit budget
pays for both $wu$ (its ordinary connector) and the closing arc (the extra
connector). This step requires no $\frac12$-unit on $wu$, which is essential when
$w\in V_{\mathcal C}$ as well and hence $wu\in M\cap E[V_{\mathcal C}]$ carries
zero matched-arc budget.

\paragraph{Attachments of whole parts.}
Suppose now that every quotient neighbor of the active part is a whole part
(either a selected piece or an intact matching cycle). A return path below may
still pass through non-adjacent ordinary singletons farther away, each of which
pays $1$ for its ordinary connector; if an untouched matching path is first
entered in this way, its first-attachment $\frac12$-unit is simply unused, and
its remaining arcs will later cross the active boundary under Rule~1.

If the active part is adjacent (in either direction) to a selected piece or to
an intact cycle of length $j\ge4$, that part has at least $2$ units available
after buying any internal arcs, which pays both its ordinary connector and the
extra connector along a simple quotient cycle through the active part.

Next, if the active part has an arc entering a vertex $u$ of an intact
$2$-cycle (digon) $\{u,v\}$, follow that entering arc and the digon arc
$uv\in M$, and use extendibility of $uv$ to return from $v$ to the active set
in $G-u$ (treating $u$ and $v$ as separate vertices and omitting $vu$). The
digon holds $2+1=3$ units, which pay for the entering arc to $u$, the arc $uv$,
and the extra connector; the symmetric reverse construction applies if an arc
leaves the digon into the active set.

It remains to treat the case where every quotient outneighbor and inneighbor of
the active part is an intact matching triangle. Let $N^+$ and
$N^-$ denote the sets of quotient outneighbors and inneighbors of the active
part. If the quotient minus the active part contains a directed path from some
$\qv c\in N^+$ to a distinct $\qv c'\in N^-$, shortcut it to a simple path:
together with the arc from the active part to $\qv c$ and the arc from $\qv c'$
back to the active part, it forms an attachment absorbing two distinct
triangles $\qv c\ne\qv c'$. Each triangle has $9/2$ units, leaving
$9/2-3-1=1/2$ unit after buying its three cycle arcs and its ordinary
connector, and the two $\frac12$-units from $\qv c$ and $\qv c'$ pay for the
extra connector.

Otherwise no path outside the active part joins any $\qv c\in N^+$ to a
distinct $\qv c'\in N^-$. Because $G$ is strongly connected, any simple return
from $\qv c\in N^+$ to the active part must end at some $\qv c'\in N^-$,
forcing $\qv c'=\qv c$; reversing the argument gives $N^+=N^-$. Fix any
$\qv c\in N^+=N^-$ representing an intact triangle $C$. We claim that input
arcs from the active set enter at least two distinct vertices of $C$. Suppose
for contradiction that all input arcs from the active set to $C$ entered a
single vertex $v\in V(C)$, and let $u$ be the predecessor of $v$ on $C$.
Extendibility of $uv\in M$ gives a directed path from $r$ to $u$ in $G-v$. Look
at the suffix of this path after its last vertex in the active set: its first
quotient vertex outside the active part lies in $N^+$. If that vertex differs
from $\qv c$, the suffix reaches $\qv c\in N^-$ outside the active part,
contradicting the absence of paths between distinct neighbors; if that vertex
is $\qv c$ itself, the first step of the suffix enters $C$ from the active set
at $v$, contradicting that the path lies in $G-v$. By directional duality, at
least two distinct vertices of $C$ also have direct input arcs into the active
set.

On a directed $3$-cycle $C$ with at least two entry vertices and at least two
exit vertices, the pigeonhole principle guarantees an entry vertex $w\in V(C)$
whose predecessor $w'$ on $C$ is an exit vertex. Enter $C$ at $w$, traverse the
two triangle arcs from $w$ around to $w'$ (visiting all three vertices of $C$
while omitting $w'w$), and exit from $w'$ back to the active set. This $4$-arc
ear absorbs $C$ at cost $4\le 9/2$.

\paragraph{Total arc count and running time.}
Summing the pre-purchased $m'$ arcs and the four budget pools gives an upper
bound of
\[
 m'+2|\mathcal C|+(n-|V_{\mathcal C}|)
       +\frac12\bigl(|M|-|M\cap E[V_{\mathcal C}]|\bigr)+\frac32(n-|M|).
\]
Using $\frac12|M|+\frac32(n-|M|)=\frac12n+(n-|M|)$ rewrites this sum as~\labelcref{eq:completion}. When $\mathcal C=\varnothing$, it simplifies to
$\frac32n+(n-|M|)\le\OPT(G)+n/2$ for a maximum matching $M\subseteq E_r$.

Testing~\labelcref{eq:root-tests} runs two graph searches in $G-u$ for each
$u\ne r$ in $O(n(n+m))$ total time. Bipartite matching runs in polynomial time,
and each attachment absorbs at least one part using standard directed graph
searches (inspecting original crossing arcs in the isolated-triangle subcase,
and breaking ties by a fixed input order).
\end{proof}

\begin{corollary}[Fractional completion]\label[corollary]{cor:fractional}
Under the hypotheses of \cref{lem:completion}, let $y\in\mathbb R_{\ge0}^E$ be
any fractional matching supported on $E_r$. Deterministic
\hyperref[proc:matching-completion]{matching completion} returns a strongly
connected spanning subgraph of $G$ retaining every selected arc of $\mathcal C$
and containing at most
\begin{equation}\label{eq:fractional}
 \frac32 n+\bigl(m'-|V_{\mathcal C}|+2|\mathcal C|\bigr)
 +\bigl(n-y(E)\bigr)-\frac12 y(E[V_{\mathcal C}])
\end{equation}
arcs.
\end{corollary}
\begin{proof}
With $\mathcal C$ fixed, rewrite~\labelcref{eq:completion} as
$\frac52n+(m'-|V_{\mathcal C}|+2|\mathcal C|)-\frac12\bigl(2|M|+|M\cap E[V_{\mathcal C}]|\bigr)$.
Because \hyperref[proc:matching-completion]{matching completion} maximizes the
linear objective $\sum_{e\in M}\bigl(2+\ind_{\{e\in E[V_{\mathcal C}]\}}\bigr)
=2|M|+|M\cap E[V_{\mathcal C}]|$ over matchings $M\subseteq E_r$ and the
bipartite fractional matching polytope is integral, the chosen matching $M$
satisfies
\[
 2|M|+|M\cap E[V_{\mathcal C}]|\ge 2y(E)+y(E[V_{\mathcal C}]).
\]
Substituting this inequality into~\labelcref{eq:completion}
yields~\labelcref{eq:fractional}.
\end{proof}

\section{Long-cycle contraction and the residual graph}\label{sec:long}

By \cref{cor:fractional}, covering both endpoints of a matched arc frees half
its weight in the completion bound. To find directed cycles whose freed
matching weight outweighs their piece excess, we solve the directed cut
relaxation~\labelcref{eq:cut-lp}. A single extreme-point optimum $x$ supplies
a fractional matching on $E_r$ and prices two spanning arborescences whose
arcs carry guaranteed $x$-weight. When $x(E)=n+o(n)$, the matching has
weight $n-o(n)$.

\begin{lemma}[A fractional matching from the cut solution]\label[lemma]{lem:cut-matching}
Given any feasible solution $x$ of~\labelcref{eq:cut-lp}, we can construct in
polynomial time a fractional matching $x'\le x$ supported on $E_r$ with
$x'(E)\ge 2n-x(E)$.
\end{lemma}
\begin{proof}
We apply a fractional version of the tree-intersection argument in
\cref{sec:completion}. For each vertex $u\ne r$, let $Z_u\subseteq V\setminus\{u\}$
be the set of vertices that reach $r$ in $G-u$. Because $r\in Z_u$ and any tail
$w\notin Z_u\cup\{u\}$ with an arc into $Z_u$ would also reach $r$ in $G-u$,
every arc in $\delta^-(Z_u)$ must have tail $u$ and therefore satisfies the
first condition of~\labelcref{eq:root-tests}. Since $\varnothing\ne Z_u\subsetneq V$,
the cut constraint gives $x(\delta^-(Z_u))\ge1$, so we may scale down $x$ on
$\delta^-(Z_u)$ (and on $\delta^+(\{r\})$ at $u=r$, where
$x(\delta^+(\{r\}))=x(\delta^-(V\setminus\{r\}))\ge1$) to obtain a vector
$a\le x$ assigning total outgoing weight $1$ to each tail $u\in V$ on arcs
satisfying the first condition of~\labelcref{eq:root-tests}. Applying the
symmetric construction in the reverse digraph gives $b\le x$ assigning total
incoming weight $1$ to each head $v\in V$ on arcs satisfying the second
condition. Setting $x'_e=\min\{a_e,b_e\}$ produces a fractional matching
$x'\le x$ supported on $E_r$. Since $\max\{a_e,b_e\}\le x_e$ for every $e\in E$,
we have
\[
 x'(E)\ge a(E)+b(E)-x(E)=2n-x(E).\qedhere
\]
\end{proof}

Compute an extreme-point optimum $x$ of~\labelcref{eq:cut-lp} and assign each
arc $e\in\supp x$ the price $p_e=1/x_e$. Compute a minimum-price spanning
in-arborescence $T^{\rm in}$ and a minimum-price spanning out-arborescence
$T^{\rm out}$ on $\supp x$, both rooted at $r$ (\cref{sec:preliminaries}).
Their union $\treeunion=T^{\rm in}\cup T^{\rm out}$ is strongly connected and
remains strongly connected under contractions. In each quotient $\widehat G$
with projection $\pi:V(G)\to V(\widehat G)$, our cycle-listing rule
(\cref{subsec:cycle-list}) inspects simple cycles of $\widehat G$ that use at
most one arc outside $F=\pi(\treeunion)$; hence buying a listed cycle of length
$j$ absorbs at least $j-1$ fresh input arcs from $\treeunion$.

\paragraph{Long-cycle contraction.}\label{proc:long-cycle-contraction}
Initialize $\widehat G=G$, $F=\treeunion$, and $\mathcal C=\varnothing$. In the
current quotient $\widehat G$, construct the cycle list of
\cref{subsec:cycle-list} and pick its first simple cycle of length at least
$\cycmin$. If no listed cycle has length at least $\cycmin$, return the
residual quotient $\widehat G$, the current family of pieces $\mathcal C$, and
$F=\pi(\treeunion)$ to \cref{sec:decomposition}.

Otherwise, select one representative input arc for each quotient arc on the
chosen cycle, taking a representative in $E(\treeunion)$ for every cycle arc
belonging to $F$. Merge these newly bought arcs with all previously selected
pieces in $\mathcal C$ whose quotient vertices lie on the cycle, forming a
single strongly connected piece on their combined preimage, and contract the
cycle in $\widehat G$. As soon as $n-|V(\widehat G)|\ge 64n/(\cycmin-1)$, run
\hyperref[proc:matching-completion]{matching completion} in $G$ on the current
pieces $\mathcal C$ and return the resulting spanning subgraph; otherwise
update $F=\pi(\treeunion)$ and repeat.

\begin{lemma}[Long-cycle contractions]\label[lemma]{lem:long}
Fix an integer $\cycmin\ge5$ and any deterministic polynomial-time rule that,
given a digraph and a designated strongly connected spanning subgraph, lists
simple cycles using at most one arc outside that subgraph.
\hyperref[proc:long-cycle-contraction]{Long-cycle contraction} runs in
deterministic polynomial time. Write $\widehat{G}$ for the current quotient and
$n_{\rm drop}=n-|V(\widehat{G})|$ for the cumulative drop in vertex count.
\begin{enumerate}
\item If $n_{\rm drop}$ reaches $64n/(\cycmin-1)$, the algorithm returns a
  strongly connected spanning subgraph of $G$ having at most
  $3n/2-n/\cycmin^2$ arcs whenever $\OPT(G)-n\le n/\cycmin^2$.
\item If every listed cycle has length less than $\cycmin$ before $n_{\rm drop}$
  reaches $64n/(\cycmin-1)$, the algorithm returns $\widehat{G}$, its
  designated spanning subgraph $F=\pi(\treeunion)$ with $|E(F)|\le 2n$, and the
  family $\mathcal C_{\rm cyc}$ of selected subgraphs (one for each
  nonsingleton preimage of $\widehat{G}$). Letting $S_{\rm cyc}$ be the union
  of their original vertex sets and $m_{\rm cyc}$ the number of selected input
  arcs, we have
\begin{equation}\label{eq:initial-bounds}
\begin{aligned}
 |S_{\rm cyc}| &\le \frac{100n}{\cycmin},\\
 m_{\rm cyc}-|S_{\rm cyc}|+2|\mathcal C_{\rm cyc}|
   &\le \frac{200n}{\cycmin^2}.
\end{aligned}
\end{equation}
\end{enumerate}
\end{lemma}
\begin{proof}
Let $x'\le x$ be the fractional matching from \cref{lem:cut-matching}. By
\cref{sec:preliminaries} and \cref{lem:sparse}, each of $T^{\rm in}$ and
$T^{\rm out}$ has price at most
$\sum_{e\in\supp x}x_e(1/x_e)=|\supp x|\le 3n-4$, so
$\sum_{e\in E(\treeunion)}(1/x_e)\le 6n$. For any subset
$A\subseteq E(\treeunion)$, Cauchy--Schwarz bounds $x(A)$ from below, and
passing from $x$ to $x'$ loses at most $(x-x')(E)\le 2(x(E)-n)$:
\begin{equation}\label{eq:reciprocal}
\begin{aligned}
 |A|^2 &\le x(A)\sum_{e\in A}\frac1{x_e}\le 6n\,x(A),\\
 x'(A) &\ge x(A)-\bigl(x(E)-x'(E)\bigr)
          \ge \frac{|A|^2}{6n}-2\bigl(x(E)-n\bigr).
\end{aligned}
\end{equation}

After $c$ contractions, write $\mathcal C_{\rm cyc}$ for the current family of
pieces, $S_{\rm cyc}$ for their vertex union, and $m_{\rm cyc}$ for the number
of selected arcs. Each nonsingleton vertex of $\widehat G$ replaces a piece
$C\in\mathcal C_{\rm cyc}$ by one vertex, so
$n_{\rm drop}=|S_{\rm cyc}|-|\mathcal C_{\rm cyc}|$. Contracting a cycle of
length $j\ge\cycmin$ buys $j$ arcs and reduces $|V(\widehat G)|$ by
$j-1\ge\cycmin-1$; summing over all $c$ contractions gives
$m_{\rm cyc}=n_{\rm drop}+c$ and $c\le n_{\rm drop}/(\cycmin-1)$. Since every
piece in $\mathcal C_{\rm cyc}$ was formed (and possibly merged) during at
least one contraction, $|\mathcal C_{\rm cyc}|\le c$, and the piece excess
satisfies
\begin{equation}\label{eq:contraction-count}
 \begin{aligned}
 m_{\rm cyc}-|S_{\rm cyc}|+2|\mathcal C_{\rm cyc}|
   &= (m_{\rm cyc}-n_{\rm drop})+|\mathcal C_{\rm cyc}|\\
   &= c+|\mathcal C_{\rm cyc}|
    \le 2c \le \frac{2n_{\rm drop}}{\cycmin-1}.
 \end{aligned}
\end{equation}
Moreover, each contracted $j$-cycle selects at least $j-1$ input arcs of
$\treeunion$, which immediately become self-loops in the next quotient and
hence are distinct across all contractions. Thus $E[S_{\rm cyc}]$ contains a
subset $A\subseteq E(\treeunion)$ of size $|A|\ge n_{\rm drop}$.

Write $k=\OPT(G)-n\ge x(E)-n$. In \cref{cor:fractional} with $y=x'$, the
matching deficit satisfies $n-x'(E)\le x(E)-n\le k$, while~\labelcref{eq:reciprocal}
gives $-\frac12 x'(E[S_{\rm cyc}])\le -\frac12 x'(A)\le -n_{\rm drop}^2/(12n)+k$.
Combining this with~\labelcref{eq:contraction-count} bounds the completed
arc count by
\begin{equation}\label{eq:long-saving}
 \frac32 n+\frac{2n_{\rm drop}}{\cycmin-1}-\frac{n_{\rm drop}^2}{12n}+2k .
\end{equation}
When $n_{\rm drop}\ge 64n/(\cycmin-1)$, we have
$\frac{2n_{\rm drop}}{\cycmin-1}\le\frac{n_{\rm drop}^2}{32n}$, so the middle
two terms sum to at most
$-\frac{5n_{\rm drop}^2}{96n}\le -\frac{n_{\rm drop}^2}{32n}\le -\frac{128n}{(\cycmin-1)^2}$.
If $k\le n/\cycmin^2$, adding $2k$ still leaves at most $3n/2-n/\cycmin^2$ arcs.

If instead all listed cycles have length less than $\cycmin$ while
$n_{\rm drop}<64n/(\cycmin-1)$, then using $\cycmin\ge5$ (so
$\frac{\cycmin}{\cycmin-1}\le\frac54$) in $|S_{\rm cyc}|=n_{\rm drop}+|\mathcal C_{\rm cyc}|\le n_{\rm drop}+c$
and in~\labelcref{eq:contraction-count} gives
\[
 |S_{\rm cyc}|
   \le \frac{\cycmin}{\cycmin-1}n_{\rm drop}
   < \frac{64\cycmin}{(\cycmin-1)^2}n
   \le \frac{100n}{\cycmin},
 \qquad
 m_{\rm cyc}-|S_{\rm cyc}|+2|\mathcal C_{\rm cyc}|
   < \frac{128n}{(\cycmin-1)^2}
   \le \frac{200n}{\cycmin^2},
\]
establishing~\labelcref{eq:initial-bounds}. Because $|E(\treeunion)|\le 2n-2$
and contraction only discards loops, $|E(F)|\le 2n$ and at most $n-1$
contractions occur, so the procedure runs in polynomial time.
\end{proof}

\section{Local structure and isolated small pieces}\label{sec:decomposition}

At each iteration of \cref{sec:long}, we build a polynomial-size list of simple
cycles in the current quotient $\widehat{G}$ from return paths inside the image
$F=\pi(\treeunion)$ of the original tree union. Whenever every listed cycle has
length less than $\cycmin$, these short cycles yield a tree decomposition of the
underlying undirected graph of $\widehat{G}$ whose bags sit inside strongly
connected subgraphs of bounded round-trip radius.

We call a quotient vertex a \emph{hub} if it either has high degree in $F$ or
represents a large contracted piece of $G$. Away from hubs, bounded round-trip
radius forces bounded original-vertex count. When at least a $\xi$-fraction of
original vertices project far from every hub, we contract a light separator band
together with periodic depth layers to isolate bounded-size far regions, whose
exact solutions beat the $3/2$ guarantee by $\Omega(\xi n)$ arcs.

For $\qv{u},\qv{v}\in V(\widehat G)$, write $\dist_{\widehat G}(\qv{u},\qv{v})$
for the shortest directed path length from $\qv{u}$ to $\qv{v}$ in
$\widehat G$, and define the \emph{round-trip distance}
\[
 d_{\rm rt}(\qv{u},\qv{v})
   =\dist_{\widehat G}(\qv{u},\qv{v})+\dist_{\widehat G}(\qv{v},\qv{u}),
\]
which is symmetric and satisfies the triangle inequality. For a subset
$\hubset\subseteq V(\widehat G)$, write
$d_{\rm rt}(\qv{v},\hubset)=\min_{\qv{h}\in\hubset}d_{\rm rt}(\qv{v},\qv{h})$
(with $d_{\rm rt}(\qv{v},\varnothing)=\infty$), so both legs of the round trip
visit the same vertex $\qv{h}\in\hubset$. We say that a strongly
connected subgraph $U\subseteq\widehat G$ containing $\qv{v}$ has
\emph{round-trip radius} at most $r_0$ about $\qv{v}$ if for every
$\qv{w}\in V(U)$, the subgraph $U$ contains a closed directed walk through
$\qv{v}$ and $\qv{w}$ of length at most $r_0$.

\begin{lemma}[Vertices far from hubs]\label[lemma]{lem:metric}
Let $\widehat{G}$ be obtained from $G$ by contracting vertex-disjoint strongly
connected subgraphs, let $\pi:V(G)\to V(\widehat{G})$ be the projection, and
let $F$ be a strongly connected spanning subgraph of $\widehat{G}$ with at most
$2n$ arcs. There is a deterministic polynomial-time procedure listing simple
cycles of $\widehat{G}$, each with at most one arc outside $F$, that has the
following property.

Suppose every listed cycle has length less than an integer $\cycmin\ge2$. For
an integer $\hubcap\ge1$, define the set of hubs by
\[
 \hubset=\bigl\{\qv{v}\in V(\widehat{G}):|\pi^{-1}(\qv{v})|>\hubcap
                     \text{ or }\deg_F(\qv{v})>\hubcap\bigr\},
\]
where $\deg_F(\qv{v})$ counts incoming plus outgoing incidences in $F$. Then
$|\hubset|\le 5n/\hubcap$. Moreover, for $0<\xi<1$, set
\begin{equation}\label{eq:radius}
 R=13\cycmin+\left\lceil\frac{512(\cycmin+1)}\xi\right\rceil+3.
\end{equation}
If at least $\xi n$ original vertices $u\in V(G)$ satisfy
$d_{\rm rt}(\pi(u),\hubset)>R$, a deterministic algorithm returns a strongly
connected spanning subgraph of $G$ with at most $3\OPT(G)/2-\xi n/4$ arcs, in
time polynomial in $n+m$ for fixed $\cycmin,\hubcap,\xi$.
\end{lemma}

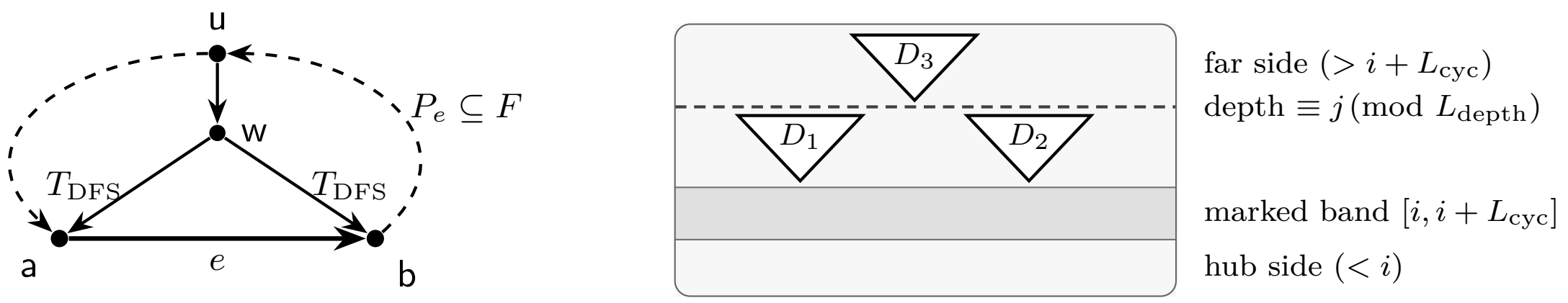
\begin{figure}[t]
\centering
\begin{tikzpicture}[>=Stealth,x=1cm,y=1cm,every node/.style={font=\small}]
  % Panel (a)
  \begin{scope}[shift={(0,0)}]
    \node[circle,fill,inner sep=1.6pt,label=above:{$\qv u$}] (u) at (3.0,1.45) {};
    \node[circle,fill,inner sep=1.5pt,label=right:{$\qv w$}] (w) at (3.0,0.72) {};
    \node[circle,fill,inner sep=1.6pt,label=below left:{$\qv a$}] (a) at (1.55,-0.25) {};
    \node[circle,fill,inner sep=1.6pt,label=below right:{$\qv b$}] (b) at (4.45,-0.25) {};

    \draw[->,thick] (u) -- (w);
    \draw[->,thick] (w) -- node[left,font=\footnotesize] {$T_{\rm DFS}$} (a);
    \draw[->,thick] (w) -- node[right,font=\footnotesize] {$T_{\rm DFS}$} (b);
    \draw[->,very thick] (a) -- node[below,font=\footnotesize] {$e$} (b);
    \draw[->,thick,dashed] (b) .. controls (5.45,0.65) and (4.35,1.45) ..
      node[right,font=\footnotesize] {$P_e\subseteq F$} (u);
    \draw[->,thick,dashed] (u) .. controls (1.65,1.45) and (0.55,0.65) .. (a);

    \node[align=center,font=\footnotesize] at (3.0,-1.25)
      {\textbf{(a)} $C_e=e+P_e$ and $T_{\rm DFS}$-descents from $\qv u$\\enclose the tree path between $\qv a$ and $\qv b$.};
  \end{scope}

  % Panel (b)
  \begin{scope}[shift={(7.0,0)}]
    % Zones
    \fill[black!3] (0.2,0.22) rectangle (4.8,1.72);
    \fill[black!12] (0.2,-0.26) rectangle (4.8,0.22);
    \fill[black!3] (0.2,-0.78) rectangle (4.8,-0.26);
    \draw[black!55] (0.2,-0.26) -- (4.8,-0.26);
    \draw[black!55] (0.2,0.22) -- (4.8,0.22);
    \draw[thick,densely dashed,black!75] (0.2,0.96) -- (4.8,0.96);
    \draw[rounded corners=4pt,black!60] (0.2,-0.78) rectangle (4.8,1.72);

    % Subtrees in far zone
    \draw[thick,fill=white] (1.35,0.28) -- (0.78,0.88) -- (1.92,0.88) -- cycle;
    \draw[thick,fill=white] (3.45,0.28) -- (2.88,0.88) -- (4.02,0.88) -- cycle;
    \draw[thick,fill=white] (2.40,1.02) -- (1.83,1.62) -- (2.97,1.62) -- cycle;
    \node[font=\scriptsize] at (1.35,0.69) {$D_1$};
    \node[font=\scriptsize] at (3.45,0.69) {$D_2$};
    \node[font=\scriptsize] at (2.40,1.43) {$D_3$};

    \node[anchor=west,font=\scriptsize] at (4.92,1.34) {far side ($>i+\cycmin$)};
    \node[anchor=west,font=\scriptsize] at (4.92,0.94) {depth $\equiv j\!\!\pmod{L_{\rm depth}}$};
    \node[anchor=west,font=\scriptsize] at (4.92,-0.02) {marked band $[i,i+\cycmin]$};
    \node[anchor=west,font=\scriptsize] at (4.92,-0.52) {hub side ($<i$)};

    \node[align=center,font=\footnotesize] at (3.6,-1.25)
      {\textbf{(b)} Distance band and periodic $T_{\rm DFS}$-depth cuts\\isolate bounded-size far pieces $D_1,D_2,D_3\in\mathcal D_{\rm ex}$.};
  \end{scope}
\end{tikzpicture}
\caption{Cycle-list enclosures on $T_{\rm DFS}$ and separator contraction away from hubs.}
\label{fig:separator-contraction}
\end{figure}

\subsection{The cycle list and bounded-radius bags}\label{subsec:cycle-list}

Run a directed depth-first search in $F$. Because $F$ is strongly connected,
its directed search tree $T_{\rm DFS}$ spans $V(\widehat{G})$.

For every quotient arc $e$ of $\widehat{G}$ from $\qv{a}$ to $\qv{b}$, choose a
simple directed return path $P_e$ in $F$ from $\qv{b}$ to $\qv{a}$ and add the
simple cycle $C_e=e+P_e$ to the list. In addition, whenever $e\in E(F)$, let
$\qv{u}$ be the vertex of $C_e$ discovered earliest by the DFS
(\cref{fig:separator-contraction}(a)). Because every other vertex of $C_e$ is
undiscovered when $\qv{u}$ is reached and is reachable from $\qv{u}$ along
$C_e$, all vertices of $C_e$ lie in the $T_{\rm DFS}$-subtree of $\qv{u}$. For
each endpoint $\qv{v}\in\{\qv{a},\qv{b}\}$, concatenate the $T_{\rm DFS}$-path
from $\qv{u}$ down to $\qv{v}$ with the segment of $C_e$ from $\qv{v}$ back to
$\qv{u}$, decompose this closed walk into simple directed cycles, and add them
to the list as well (these extra cycles lie wholly in $F$). The contraction
loop of \cref{lem:long} tests this polynomial-size list for a cycle of length
at least $\cycmin$.

Assume henceforth that every listed cycle has length less than $\cycmin$. Since
$T_{\rm DFS}$ is acyclic, every simple cycle in the decomposition of the closed
walk for $\qv{v}\in\{\qv{a},\qv{b}\}$ must meet $C_e$. Consequently, within the
strongly connected union of $C_e$ and the two $T_{\rm DFS}$-descents from
$\qv{u}$ to $\qv{a}$ and $\qv{b}$, every descent vertex can reach $C_e$ and
return in fewer than $\cycmin$ arcs, and hence has round-trip walk length less
than $2\cycmin$ through $\qv{u}$. Moreover, because $\qv{u}$ is a common
ancestor of $\qv{a}$ and $\qv{b}$ in $T_{\rm DFS}$, their lowest common
ancestor is a descendant of $\qv{u}$, so the two descents contain the unique
undirected path between $\qv{a}$ and $\qv{b}$ in $T_{\rm DFS}$.

Recall that a tree decomposition of the underlying undirected graph of
$\widehat{G}$ assigns a vertex subset (\emph{bag}) to each node of a tree so
that every vertex appears in a bag, both endpoints of every arc appear together
in some bag, and the bags containing any fixed vertex form a connected subtree.
Index the bags by the vertices of $T_{\rm DFS}$, starting each \emph{base bag}
at $\qv{v}$ with $\{\qv{v}\}$. For every arc $e\in E(F)$, insert both endpoints
of $e$ into every base bag along the undirected $T_{\rm DFS}$-path joining
them, and attach to each node $\qv{v}$ on that path the strongly connected
union of $C_e$ and the two descents constructed above. These base bags already
form a tree decomposition of $F$. Uniting $\{\qv{v}\}$ with the attached
subgraphs over all $F$-arcs whose tree paths pass through $\qv{v}$ gives a
strongly connected base set containing the base bag of $\qv{v}$: since $\qv{v}$
has round-trip walk length less than $2\cycmin$ to each corresponding cycle
root $\qv{u}$ and $\qv{u}$ has round-trip walk length less than $2\cycmin$ to
every vertex of its subgraph, this base set has round-trip radius at most
$4\cycmin$ about $\qv{v}$.

Finally, to cover arcs outside $F$, keep the base bags fixed as a reference and
define enlarged bags $B_{\qv{v}}$ and strongly connected \emph{enclosures}
$U_{\qv{v}}$ as follows: for every arc $e$ of $\widehat{G}$ and every node
$\qv{v}$ whose \emph{base} bag contains at least one vertex of $P_e$, add both
endpoints of $e$ to $B_{\qv{v}}$ and unite $C_e$ with $U_{\qv{v}}$. Because
consecutive vertices of $P_e\subseteq F$ share a base bag, the nodes whose base
bags meet $P_e$ form a connected subtree of $T_{\rm DFS}$ that includes every
base bag containing either endpoint of $e$; hence $(B_{\qv{v}})_{\qv{v}\in V(T_{\rm DFS})}$
is a tree decomposition of the underlying undirected graph of $\widehat{G}$.
Any vertex on an added cycle $C_e$ can be visited from $\qv{v}$ by detours of
fewer than $\cycmin$ arcs off a closed walk in the base set, so $U_{\qv{v}}$
has round-trip radius at most $5\cycmin$ about $\qv{v}$. Note also that
adjacent nodes of $T_{\rm DFS}$ satisfy $d_{\rm rt}(\qv{u},\qv{v})<\cycmin$, as
their tree arc $e\in E(F)$ lies on $C_e$.

\subsection{Bag size and overlap away from hubs}

Pulling $(B_{\qv{v}})$ back along $\pi$ gives expanded bags $\pi^{-1}(B_{\qv{v}})$
that form a tree decomposition of the underlying undirected graph of $G$
(covering both internal piece arcs and crossing arcs), while each preimage
$\pi^{-1}(U_{\qv{v}})$ is strongly connected in $G$ via the selected internal
arcs of the contracted pieces.

Because $\sum_{\qv{v}}|\pi^{-1}(\qv{v})|=n$ and
$\sum_{\qv{v}}\deg_F(\qv{v})=2|E(F)|\le 4n$, Markov's inequality gives
$|\hubset|\le 5n/\hubcap$. Now suppose $d_{\rm rt}(\qv{v},\hubset)>6\cycmin$.
For any $\qv{w}\in V(U_{\qv{v}})$, take a closed walk through $\qv{v}$ and
$\qv{w}$ in $U_{\qv{v}}$ of length at most $5\cycmin$. Traverse this walk in
reverse order: each arc $e$ from $\qv{a}$ to $\qv{b}$ becomes a backward step
from $\qv{b}$ to $\qv{a}$, which we replace by its forward return path
$P_e\subseteq F$ of length less than $\cycmin$. The result is a closed directed
walk in $F$ through $\qv{v}$ and $\qv{w}$ of length at most $5\cycmin^2$. Every
vertex $\qv{z}$ along this $F$-walk lies on a listed cycle $C_e$ intersecting
the original $5\cycmin$-walk, so
$d_{\rm rt}(\qv{v},\qv{z})<5\cycmin+\cycmin=6\cycmin$; because
$d_{\rm rt}(\qv{v},\hubset)>6\cycmin$, the triangle inequality gives
$\qv{z}\notin\hubset$, meaning $\deg_F(\qv{z})\le\hubcap$ and
$|\pi^{-1}(\qv{z})|\le\hubcap$.

Starting from any fixed quotient vertex, forward walks in $F$ of length at most
$5\cycmin^2$ whose vertices all have outdegree at most $\hubcap$ in $F$ can
reach at most
\[
 \sum_{j=0}^{5\cycmin^2}\hubcap^j
 \le (\hubcap+1)^{5\cycmin^2+1}
 \eqqcolon n_{\rm loc}
\]
quotient vertices. Applying this count from a center $\qv{v}$ with
$d_{\rm rt}(\qv{v},\hubset)>6\cycmin$ bounds $|V(U_{\qv{v}})|\le n_{\rm loc}$
and therefore $|\pi^{-1}(U_{\qv{v}})|\le\hubcap\,n_{\rm loc}$. Applying the
same count starting from $\pi(u)$ for any original vertex $u\in V(G)$ shows
that $u$ belongs to $\pi^{-1}(U_{\qv{v}})$ for at most $n_{\rm loc}$ centers
$\qv{v}$ satisfying $d_{\rm rt}(\qv{v},\hubset)>6\cycmin$.

\subsection{Separator contraction and local solutions}\label{subsec:local-solutions}

Even when individual bags are small, adjacent bags may share multiple vertices.
To solve far regions independently, \emph{separator contraction} marks a light
band of boundary bags around $\hubset$ together with periodic DFS-depth layers
on the far side of the band, and contracts their enclosures in $G$ into single
attachment vertices (\cref{fig:separator-contraction}(b)).

Specifically, for each of the $K=\lceil 512(\cycmin+1)/\xi\rceil$ integers
$i\in\{6\cycmin+1,\dots,6\cycmin+K\}$, band $i$ marks all nodes $\qv{v}$ whose
centers satisfy $i\le d_{\rm rt}(\qv{v},\hubset)\le i+\cycmin$. Choose $i$
minimizing the number of original vertices in
$A_i=\bigcup_{i\le d_{\rm rt}(\qv{v},\hubset)\le i+\cycmin}\pi^{-1}(U_{\qv{v}})$.
Call a node $\qv{v}$ \emph{on the hub side} of band $i$ if
$d_{\rm rt}(\qv{v},\hubset)<i$, and \emph{on the far side} if
$d_{\rm rt}(\qv{v},\hubset)>i+\cycmin$. Measuring DFS depth in $T_{\rm DFS}$ as
tree distance from the root, set $L_{\rm depth}=\lceil 32n_{\rm loc}/\xi\rceil$,
and for each residue class $j\in\{0,\dots,L_{\rm depth}-1\}$ let $D_j$ be the
union of $\pi^{-1}(U_{\qv{v}})$ over nodes $\qv{v}$ on the far side of band $i$
whose DFS depth is congruent to $j\pmod{L_{\rm depth}}$. Choose $j$ minimizing
$|D_j|$ and mark those nodes as well, setting $Z=A_i\cup D_j$ for the chosen
pair $(i,j)$.

If $u\in A_{i'}$, then $u\in\pi^{-1}(U_{\qv{v}})$ for some $\qv{v}$ with
$i'\le d_{\rm rt}(\qv{v},\hubset)\le i'+\cycmin$; since $U_{\qv{v}}$ has
round-trip radius at most $5\cycmin$ about $\qv{v}$, the triangle inequality
restricts $d_{\rm rt}(\pi(u),\hubset)$ to $[i'-5\cycmin,\,i'+6\cycmin]$, so
each $u\in V(G)$ belongs to $A_{i'}$ for at most $11\cycmin+1$ indices $i'$.
Likewise, each $u\in V(G)$ belongs to $\pi^{-1}(U_{\qv{v}})$ for at most
$n_{\rm loc}$ far-side centers $\qv{v}$, and hence to at most $n_{\rm loc}$
sets $D_j$. Averaging over $i$ and then over $j$ gives
\begin{equation}\label{eq:separator-cost}
 \begin{aligned}
 |Z| &\le \frac{1}{K}\sum_{i'} |A_{i'}|
          +\frac{1}{L_{\rm depth}}\sum_{j=0}^{L_{\rm depth}-1} |D_j|\\
     &\le \frac{(11\cycmin+1)n}{K}
          +\frac{n_{\rm loc}\,n}{L_{\rm depth}}
      \le \frac{\xi n}{32}+\frac{\xi n}{32}
      = \frac{\xi n}{16}.
 \end{aligned}
\end{equation}

Each marked node $\qv{v}$ carries the strongly connected subgraph
$\pi^{-1}(U_{\qv{v}})$ of $G$. Merge intersecting marked subgraphs until they
are pairwise vertex-disjoint (their vertex union remains $Z$), and contract
each merged subgraph in $G$ to obtain a quotient $G'$ with projection
$\sigma:V(G)\to V(G')$. Every vertex $u\in V(G)\setminus Z$ stays a singleton
preimage $\sigma^{-1}(\sigma(u))=\{u\}$, while lifting any spanning subgraph of
$G'$ back to $G$ requires adding at most one in-arborescence and one
out-arborescence inside each contracted component of $Z$, costing at most
$2|Z|$ arcs.

Projecting the expanded bags gives $\sigma(\pi^{-1}(B_{\qv{v}}))$; because the
bags meeting any connected contracted subgraph induce a connected subtree of
$T_{\rm DFS}$, these projected bags form a tree decomposition of the underlying
undirected graph of $G'$, and every marked node $\qv{v}$ now has a
\emph{singleton} bag $\{\qv{z}\}\subseteq\sigma(Z)$ (since
$B_{\qv{v}}\subseteq V(U_{\qv{v}})$). Group each connected component of
unmarked nodes of $T_{\rm DFS}$ into one piece, give each marked node its own
$1$-vertex piece, and contract these node groups in $T_{\rm DFS}$ to form the
\emph{piece tree}, where each piece $D$ carries the induced subgraph of $G'$ on
the union of its bags.

Across every edge of the piece tree, at least one endpoint is a marked
$1$-vertex bag $\{\qv{z}\}$, so the subtrees on the two sides of that edge
share at most the single vertex $\qv{z}$ (by connected bag occurrences) and no
arc of $G'$ joins their interiors (by arc coverage); if both sides contain
vertices outside $\{\qv{z}\}$, strong connectivity of $G'$ forces them to share
$\qv{z}$. Whenever $G'$ is decomposed into two induced subgraphs meeting only
at $\qv{z}$ with no crossing interior arcs, no simple directed path with both
endpoints on one side can visit the other side without repeating $\qv{z}$.
Every strongly connected spanning subgraph of $G'$ therefore restricts to a
strongly connected spanning subgraph on each side, and conversely the arc-disjoint
union of strongly connected spanning subgraphs of the two sides is strongly
connected on $G'$. Peeling leaves of the piece tree inductively shows that
every piece $D$ is strongly connected and satisfies
\begin{equation}\label{eq:piece-optima}
 \OPT(G')=\sum_{D\text{ a piece}}\OPT(D)\le \OPT(G),
\end{equation}
where $1$-vertex pieces contribute $0$ and the inequality follows by projecting
an optimal SCSS of $G$ onto $G'$.

Furthermore, every unmarked component of $T_{\rm DFS}$ lies either wholly on
the hub side or wholly on the far side of band $i$: if $\qv{v}$ is on the hub
side ($d_{\rm rt}(\qv{v},\hubset)\le i-1$), every $T_{\rm DFS}$-neighbor
$\qv{w}$ satisfies $d_{\rm rt}(\qv{w},\hubset)<(i-1)+\cycmin=i+\cycmin-1\le i+\cycmin$,
so $\qv{w}$ cannot jump to the far side without being marked in the band. On
the far side, marking one residue class modulo $L_{\rm depth}$ restricts each
unmarked component to fewer than $L_{\rm depth}$ consecutive DFS depths, and
every node there has outdegree at most $\hubcap$ in $F\supseteq T_{\rm DFS}$.
Such a component therefore contains at most $(\hubcap+1)^{L_{\rm depth}}$ bags,
each with $|\pi^{-1}(B_{\qv{v}})|\le\hubcap\,n_{\rm loc}$ vertices, bounding
every far-side piece $D$ by
\begin{equation}\label{eq:piece-size}
 |V(D)|\le\hubcap\,n_{\rm loc}(\hubcap+1)^{L_{\rm depth}}.
\end{equation}

We can now solve each piece $D$: if $D$ comes from an unmarked component on the
far side of the band, keep at most one arc per ordered pair of distinct
vertices and find an exact minimum SCSS of $D$ in $O(1)$ time (for fixed
$\cycmin,\hubcap,\xi$) by subset enumeration; on every other piece $D$ with
$|V(D)|\ge2$, run \hyperref[proc:matching-completion]{matching completion} in
$D$ with $\mathcal C=\varnothing$, obtaining at most $\frac32\OPT(D)$ arcs by
\cref{lem:completion}. Lift the union of all piece solutions back to $G$ at
cost at most $2|Z|$.

Suppose at least $\xi n$ original vertices $u$ satisfy
$d_{\rm rt}(\pi(u),\hubset)>R$. At least $\xi n-|Z|$ of them lie outside $Z$;
their projections $\sigma(u)$ are distinct vertices of $G'\setminus\sigma(Z)$,
so each belongs to a unique piece $D$. Because $i\le 6\cycmin+K$ while
$R>6\cycmin+K+\cycmin\ge i+\cycmin$ by~\labelcref{eq:radius}, the center
$\pi(u)$ lies on the far side of band $i$, placing $\sigma(u)$ in an exactly
solved piece $D\in\mathcal D_{\rm ex}$ together with at least one outgoing arc
of $\sigma(u)$ in $G'$. Summing these distinct outgoing arcs gives
$\sum_{D\in\mathcal D_{\rm ex}}\OPT(D)\ge\xi n-|Z|$. Combining~\labelcref{eq:separator-cost,eq:piece-optima}
bounds the lifted output $E_{\rm out}$ by
\begin{equation}\label{eq:far-piece-cost}
 \begin{aligned}
 |E_{\rm out}|
 &\le \sum_{D\in\mathcal D_{\rm ex}}\OPT(D)
      +\frac32\sum_{D\notin\mathcal D_{\rm ex}}\OPT(D)+2|Z|\\
 &= \frac32\OPT(G')-\frac12\sum_{D\in\mathcal D_{\rm ex}}\OPT(D)+2|Z|\\
 &\le \frac32\OPT(G)-\frac{\xi n}{2}+\frac52|Z|
  \le \frac32\OPT(G)-\left(\frac12-\frac5{32}\right)\xi n
  \le \frac32\OPT(G)-\frac{\xi n}{4},
 \end{aligned}
\end{equation}
completing the proof of \cref{lem:metric}.

\section{Shared returns for paths with endpoints near hubs}\label{sec:near-hubs}

Suppose \hyperref[proc:long-cycle-contraction]{long-cycle contraction} runs out
of listed cycles of length at least $\cycmin$ before reaching the vertex-drop
threshold of \cref{lem:long}. Keep the residual quotient $\widehat{G}$, its
projection $\pi:V(G)\to V(\widehat{G})$, and the family $\mathcal C_{\rm cyc}$
of pieces selected so far; every arc of $\widehat{G}$ now has a listed return
path of length less than $\cycmin$ inside $F=\pi(\treeunion)$.

We work at two block sizes, $\shortsize$ and $\longsize$ vertices, with
$2\le\shortsize<\longsize$. For either size $s\in\{\shortsize,\longsize\}$, let
$R_s$ be the round-trip radius given by~\labelcref{eq:radius} for a prescribed
far-vertex fraction $\xi_s\in(0,1)$; to keep subscripts compact, we write
$\mathrm{short}$ and $\mathrm{long}$ in place of $\shortsize$ and $\longsize$
on quantities indexed by $s$ (such as $R_s,\xi_s,b_s,b_s^{(1)},\rho_s,Z_s,\omega_s$).
If at least $\xi_s n$ original vertices $u$ satisfy
$d_{\rm rt}(\pi(u),\hubset)>R_s$ for either $s\in\{\shortsize,\longsize\}$,
separator contraction already returns the improved solution of
\cref{lem:metric}. Assume henceforth that fewer than $\xi_s n$ original
vertices are far at each size. Then nearly all vertices project near $\hubset$,
and buying a single network of routes between nearby hubs lets many paths share
the middle of their return.

\subsection{The hub forest}\label{subsec:shared-returns}

To close paths whose endpoints project near hubs, form an auxiliary undirected
graph on $\hubset$ joining any two hubs $\qv{h},\qv{h}'$ with
$d_{\rm rt}(\qv{h},\qv{h}')\le
\joinradius=2\max(R_{\mathrm{short}},R_{\mathrm{long}})+\cycmin\longsize$, and
choose a spanning forest of this graph. For each forest edge
$\{\qv{h},\qv{h}'\}$, select a shortest directed path from $\qv{h}$ to
$\qv{h}'$ and one from $\qv{h}'$ to $\qv{h}$ in $\widehat{G}$, lifting them to
$G$ by choosing one input arc per quotient arc and relying on previously
selected internal arcs inside nonsingleton preimages. Include the preimage of
every isolated hub as well (adding a $1$-vertex piece if its preimage is a
singleton), combine these subgraphs with $\mathcal C_{\rm cyc}$, and merge any
intersecting subgraphs until they form a pairwise disjoint family of initial
pieces serving both block sizes.

Write $U\subseteq V(G)$ for the vertex union of these initial pieces and
$\initexcess$ for their piece excess (number of selected input arcs, minus
$|U|$, plus twice the number of pieces). By \cref{lem:metric}, there are
$|\hubset|\le 5n/\hubcap$ hubs. The spanning forest has fewer than $|\hubset|$
edges, each buying at most $\joinradius$ quotient arcs, and every nonsingleton
preimage met along these routes already lies in $S_{\rm cyc}$, so at most
$\joinradius|\hubset|$ new singleton vertices from forest routes plus at most
$|\hubset|$ isolated singleton hubs enter $U\setminus S_{\rm cyc}$. Before
merging with $\mathcal C_{\rm cyc}$, the hub forest forms at most $|\hubset|$
connected components (contributing at most $2|\hubset|$ to the attachment term,
which merging can only decrease). Consequently,
\begin{align}
 |U|&\le |S_{\rm cyc}|+(\joinradius+1)\frac{5n}{\hubcap},\nonumber\\
 \initexcess&\le m_{\rm cyc}-|S_{\rm cyc}|+2|\mathcal C_{\rm cyc}|
                       +(\joinradius+2)\frac{5n}{\hubcap}.
 \label{eq:merge}
\end{align}

Now consider any directed path $P$ on $s\le\longsize$ original vertices whose
endpoints $u$ and $v$ satisfy $d_{\rm rt}(\pi(u),\qv{h}_u)\le R_s$ and
$d_{\rm rt}(\pi(v),\qv{h}_v)\le R_s$ for hubs $\qv{h}_u,\qv{h}_v\in\hubset$. Each of the at most $s-1$ non-loop quotient arcs
along $\pi(P)$ has a return path of length less than $\cycmin$, so its
endpoints have round-trip distance at most $\cycmin$ in $\widehat{G}$; the
triangle inequality therefore gives
\[
 d_{\rm rt}(\qv{h}_u,\qv{h}_v)\le 2R_s+\cycmin(s-1)\le\joinradius.
\]
Hence $\qv{h}_u$ and $\qv{h}_v$ lie in the same connected component of the hub
forest. Routing in $\widehat{G}$ from $\pi(v)$ to $\qv{h}_v$, traversing the
already-purchased hub-forest piece to $\qv{h}_u$, and routing from $\qv{h}_u$
back to $\pi(u)$ closes $P$ into a directed walk in $G$ while buying at most
$d_{\rm rt}(\pi(v),\qv{h}_v)+d_{\rm rt}(\qv{h}_u,\pi(u))\le 2R_s$ new input
arcs.

\subsection{Long and short path covers}\label{subsec:path-covers}

To cover $V$ by hub-closable paths, we allow candidate paths to overlap
fractionally. A \emph{fractional path cover} assigns nonnegative weights to
simple paths of size $s$ and to $1$-vertex \emph{singleton} blocks so that
every vertex receives total covering weight $1$.

Let $\mathcal P$ be the family of all simple directed paths on $\longsize$
vertices in $G$ whose arcs belong to $E_r$ and whose endpoints $u,v$ satisfy
$d_{\rm rt}(\pi(u),\hubset)\le R_{\mathrm{long}}$ and
$d_{\rm rt}(\pi(v),\hubset)\le R_{\mathrm{long}}$. Introducing a weight
$z_P\ge0$ for each $P\in\mathcal P$ and a singleton weight $z_v\ge0$ for each
$v\in V$, solve the explicit linear program
\begin{equation}\label{eq:path-lp}
 \begin{aligned}
 \text{minimize}\quad &\sum_{v\in V}z_v\\
 \text{subject to}\quad &z_v+\sum_{P\in\mathcal P:\,v\in V(P)}z_P=1
             &&(v\in V),\\
 &z_P,z_v\ge0.
 \end{aligned}
\end{equation}
Setting $z_v=1$ and $z_P=0$ shows feasibility, and $|\mathcal P|\le n^{\longsize}$
is polynomial for fixed $\longsize$. To construct a companion short cover whose
arc-usage weights never exceed those of the long cover, we subdivide each
positive-weight path $P=(v_0,\dots,v_{\longsize-1})$ at all $\shortsize$
modulo-$\shortsize$ \emph{offsets}: for an offset
$t\in\{0,\dots,\shortsize-1\}$, cut $P$ immediately before every index
$q\equiv t\pmod{\shortsize}$, partitioning $V(P)$ into consecutive blocks of
$\shortsize$ vertices together with at most two shorter end blocks.

\begin{procedure}{Fractional path covers}\label{alg:covers}
\textbf{Input.} $G$, its quotient $\widehat G$ with $\hubset$, $E_r$, and block
sizes $\shortsize,\longsize$ with radii $R_{\mathrm{short}},R_{\mathrm{long}}$.\par
\textbf{Output.} Weighted long and short covers, and a fractional matching $y$
supported on $E_r$.
\begin{enumerate}
\item Enumerate $\mathcal P$ and solve~\labelcref{eq:path-lp}. Its
  positive-weight paths and singletons form the long cover. For each arc
  $e\in E$, set $y_e=\sum_{P\in\mathcal P:\,e\in E(P)}z_P$.
\item Initialize the short cover with singleton weights $z_v$ ($v\in V$). For
  each $P\in\mathcal P$ with $z_P>0$ and each offset
  $t\in\{0,\dots,\shortsize-1\}$, partition $P$ at offset $t$: give weight
  $z_P/\shortsize$ to each full $\shortsize$-vertex subpath whose two endpoints
  both project within round-trip distance $R_{\mathrm{short}}$ of $\hubset$,
  and replace every other subpath (including incomplete end blocks) by its
  individual vertices as singletons of weight $z_P/\shortsize$ each. Sum the
  weights of identical blocks.
\end{enumerate}
\end{procedure}

Because every offset partitions $V(P)$, averaging over the $\shortsize$ offsets
preserves total covering weight $1$ at every vertex. For
$s\in\{\shortsize,\longsize\}$, write $b_s^{(1)}$ for the total singleton
weight in the size-$s$ cover.

\begin{lemma}[Singleton weight]\label[lemma]{lem:path-cover}
If $\OPT(G)=n+k$, the covers computed by \cref{alg:covers} satisfy
\begin{equation}\label{eq:singleton-bounds}
 \begin{aligned}
 b_{\mathrm{long}}^{(1)}&\le 2\longsize(k+1)+2\xi_{\mathrm{long}} n,\\
 b_{\mathrm{short}}^{(1)}&\le b_{\mathrm{long}}^{(1)}
                  +\frac{2\shortsize n}{\longsize}+2\xi_{\mathrm{short}} n .
 \end{aligned}
\end{equation}
\end{lemma}
\begin{proof}
Inside an optimal solution of size $n+k$, take a spanning in-arborescence
rooted at $r$ plus one outgoing arc at $r$ ($n$ arcs) and a spanning
out-arborescence rooted at $r$ ($n-1$ arcs, omitting the incoming root arc used
in \cref{sec:completion}). Their intersection has at least
$n+(n-1)-(n+k)=n-k-1$ arcs, lies in $E_r$, and is acyclic because it is
contained in the out-arborescence; hence its weakly connected components
partition $V$ into at most $k+1$ directed paths (including isolated vertices).

Partition each of these paths into blocks of $\longsize$ vertices at a
uniformly random offset in $\{0,\dots,\longsize-1\}$. On each path, the two
incomplete end blocks contain fewer than $2\longsize$ vertices in total. Any
far vertex $u$ (with $d_{\rm rt}(\pi(u),\hubset)>R_{\mathrm{long}}$) serves as
the start or end vertex of a full $\longsize$-vertex block for at most $2$ of
the $\longsize$ offsets; discarding such a block turns $\longsize$ vertices
into singletons, charging at most $\frac{2}{\longsize}\cdot\longsize=2$
singletons in expectation to each of the fewer than $\xi_{\mathrm{long}}n$ far
vertices. Averaging over all $\longsize$ offsets yields a feasible solution
of~\labelcref{eq:path-lp} and proves the first inequality.

Applying the identical offset count in Step~2 of \cref{alg:covers} proves the
second inequality: the positive-weight long paths have total weight at most
$n/\longsize$, losing at most $2\shortsize(n/\longsize)$ weight to incomplete
ends, while each of the fewer than $\xi_{\mathrm{short}}n$ vertices beyond
$R_{\mathrm{short}}$ lies on long paths of combined weight at most $1$ and
accounts for at most $2$ units of singleton weight when full short blocks
ending at it are dissolved.
\end{proof}

For $s\in\{\shortsize,\longsize\}$, let $b_s$ denote the total block weight in
the size-$s$ cover (summing over both $s$-vertex paths and singletons). Because
each $s$-vertex path covers $s$ vertices while each singleton covers $1$,
summing vertex weights gives $n=b_s^{(1)}+s(b_s-b_s^{(1)})$, or
$b_s=b_s^{(1)}+(n-b_s^{(1)})/s$. Moreover, at every vertex $u\in V$, at most
one arc of any simple path $P$ enters $u$ and at most one leaves $u$, so
$y(\delta^\pm(u))\le\sum_{P\ni u}z_P\le1$ and $y$ is a fractional matching on
$E_r$. Every block in the long cover (whether an $\longsize$-vertex path or a
singleton) contains one fewer arc than vertices, so weighting by $z$ gives the
exact matching deficit $n-y(E)=b_{\mathrm{long}}$. Finally, at each offset
$t\in\{0,\dots,\shortsize-1\}$, an arc $e\in E(P)$ belongs to at most one full
short block of $P$, so the total weight of short cover paths containing $e$ is
at most $y_e$.

\subsection{Shared-return rounding}\label{subsec:block-construction}

To convert a fractional cover of size $s\in\{\shortsize,\longsize\}$ into
discrete pieces for matching completion, we sample $N_s=\lfloor b_s/2\rfloor$
blocks independently, choosing each block of weight $z$ with probability
$z/b_s$, and close the sampled paths through the hub forest (\emph{shared-return
rounding}). In our deterministic algorithm, \cref{subsec:derandomize} evaluates
a closed-form upper bound on the expected final arc count at both sizes, picks
the better size $s$, fixes the $N_s$ blocks sequentially by conditional
expectations, and passes that deterministic block sequence to
\cref{alg:rounding}.

\begin{procedure}{Shared-return rounding for a block sequence}\label{alg:rounding}
\textbf{Input.} $G$, $E_r$, the initial pieces on $U$ built from
$\mathcal C_{\rm cyc}$ and the hub forest, a sequence of $N_s$ blocks from the
size-$s$ cover, and the endpoint-to-hub routes with their input-arc
representatives.\par
\textbf{Output.} A strongly connected spanning subgraph of $G$.
\begin{enumerate}
\item Process the $N_s$ blocks in order. If the block is an $s$-vertex path
  $P$, close $P$ via its endpoint hubs as in \cref{subsec:shared-returns} and
  lift the closed walk to $G$. Form the union of this lifted closed walk with
  all previously selected pieces it intersects, contract each of those pieces to
  a single vertex, and choose a spanning in-arborescence and a spanning
  out-arborescence rooted at a common vertex in the resulting strongly
  connected quotient; buy the input-arc representatives of both arborescences
  and merge all met pieces with the walk into one piece. If the block is a
  singleton $\{v\}$ and $v$ does not yet belong to any selected piece, add
  $(\{v\},\varnothing)$ as a $1$-vertex piece.
\item Run \hyperref[proc:matching-completion]{matching completion} in $G$ on
  the resulting disjoint family $\mathcal C$ (selecting $M\subseteq E_r$ to
  maximize $2|M|+|M\cap E[V_{\mathcal C}]|$) and return its spanning subgraph.
\end{enumerate}
\end{procedure}

\section{Final analysis}\label{sec:final-analysis}

We now analyze the near-hub candidate produced by \cref{alg:rounding}, compare
the expected guarantees at block sizes $\shortsize$ and $\longsize$,
derandomize the block sequence via conditional expectations, and fix the
constants proving \cref{thm:main}.

\subsection{Piece excess and matching weight covered by one block}\label{subsec:draw-cost}

Fix $s\in\{\shortsize,\longsize\}$ and assume $b_s\ge64$ (verified for our
parameter choices in \cref{sec:parameters}). Because every vertex has unit
covering weight, a single draw hits a given vertex with probability
$\alpha=1/b_s$, and $N_s=\lfloor b_s/2\rfloor$ independent draws hit it with
probability $\beta=1-(1-\alpha)^{N_s}$. Since $b_s\ge64$, we have
$1/2-1/64\le N_s\alpha\le 1/2$, whence $1/3\le N_s\alpha\le 1/2$ and
\[
 \frac38 < 1-e^{-N_s\alpha}\le \beta \le N_s\alpha \le \frac12.
\]
Let $X_i$ denote the union of the vertex sets of the first $i$ drawn blocks, and
put $S_i=U\cup X_i$ (note that vertices visited only along lifted endpoint
routes may already lie in the selected pieces $V_{\mathcal C_i}$, so
$S_i\subseteq V_{\mathcal C_i}$ throughout).

Consider the $(i+1)$-st draw in Step~1 of \cref{alg:rounding}, and write
$m'_i,\mathcal C_i$ for the selected arc count and piece family after $i$
draws. If the $(i+1)$-st block $B$ is an $s$-vertex path $P$, its lifted closed
walk visits at least one hub and therefore meets $j\ge1$ pieces in
$\mathcal C_i$. Contracting those $j$ pieces in the union of the walk and the
met pieces leaves a strongly connected graph in which each of the two spanning
arborescences induces a forest on the $j$ contracted piece vertices, using at
most $j-1$ arcs between them ($2(j-1)$ arcs in total across both arborescences).
Every other arc in the union of the two arborescences is bought at most once and
is either an arc of $E(P)\setminus E[S_i]$ (since both endpoints of any arc in
$E(P)\cap E[S_i]\subseteq E[V_{\mathcal C_i}]$ already lie in contracted pieces)
or one of the at most $2R_s$ new endpoint-route arcs. Because merging $j$
pieces into $1$ reduces $2|\mathcal C|$ by $2(j-1)$, we obtain
\[
 (m'_{i+1}-|S_{i+1}|+2|\mathcal C_{i+1}|)-(m'_i-|S_i|+2|\mathcal C_i|)
 \le |E(P)\setminus E[S_i]|-|V(P)\setminus S_i|+2R_s.
\]
If instead $B=\{v\}$ is a singleton block, adding it increases $2|\mathcal C|$
by at most $2$ while $|S_{i+1}|-|S_i|=\ind_{\{v\notin S_i\}}$, so its increase
is at most $-\ind_{\{v\notin S_i\}}+2$. Define the $(i+1)$-st \emph{charge} to
be $|E(P)\setminus E[S_i]|-|V(P)\setminus S_i|+2R_s$ for a path $P$ and
$-\ind_{\{v\notin S_i\}}+2$ for a singleton $\{v\}$, and write $A$ for the sum
of all $N_s$ charges. Since $S_0=U$ and $S_{N_s}\subseteq V_{\mathcal C}$ at
termination, summing over $i=0,\dots,N_s-1$ gives
$m'-|V_{\mathcal C}|+2|\mathcal C|\le\initexcess+A$.

Condition on the first $i$ draws (fixing $S_i$). Along a drawn path $P$, break
$V(P)\setminus S_i$ into maximal contiguous runs of fresh vertices: a run of $q$ fresh vertices bounded on both
sides by vertices of $S_i$ contributes $q+1$ arcs to $E(P)\setminus E[S_i]$,
subtracts $q$ fresh vertices, and crosses $\delta(S_i)$ twice (net $+1$, equal
to half its boundary crossings), whereas a run reaching either endpoint of $P$
contributes at most $q$ arcs minus $q$ vertices (net $\le 0$). Thus, for both
paths and singletons, the block's charge before adding $2R_s$ or $2$ is at most
$\frac12|E(B)\cap\delta(S_i)|$. Because the size-$s$ cover uses each arc $e$ to
total weight at most $y_e$ (\cref{subsec:path-covers}), a single draw contains
$e$ with probability at most $\alpha y_e$, so this boundary portion of the
expected charge is at most $\frac12\alpha\,y(\delta(S_i))$.

For $u,v\in V$, let $\rho_s(u,v)$ denote the conditional probability that a
single draw from the size-$s$ cover contains $v$ given that it contains $u$;
since $u$ has total covering weight $1$, $\rho_s(u,v)$ equals the total weight
of size-$s$ blocks containing both $u$ and $v$ (with $\rho_s(u,u)=1$), and a
single draw hits both $u$ and $v$ with probability $\alpha\rho_s(u,v)$. An arc
$uv\in\delta(S_i)$ enters $E[S_{i+1}]$ whenever its unique endpoint outside
$S_i$ is hit (probability $\alpha$), while $uv\in E[V\setminus S_i]$ enters
$E[S_{i+1}]$ on a joint hit (probability $\alpha\rho_s(u,v)$):
\[
 \E\bigl[y(E[S_{i+1}])-y(E[S_i])\;\big|\;S_i\bigr]
 = \alpha\,y(\delta(S_i))
   +\alpha\sum_{uv\in E[V\setminus S_i]} y_{uv}\rho_s(u,v).
\]
Subtracting half of this expected increase from the expected $(i+1)$-st charge
cancels $\frac12\alpha\,y(\delta(S_i))$. Setting
$Z_s=\sum_{uv\in E[V\setminus U]}y_{uv}\rho_s(u,v)$, union bound shows that each
$uv\in E[V\setminus U]$ still lies in $E[V\setminus S_i]$ with probability at
least $1-2i\alpha$, and summing over $0\le i<N_s$ gives
$\sum_{i=0}^{N_s-1}\alpha(1-2i\alpha)\ge N_s\alpha(1-N_s\alpha)\ge \frac29>\frac16$.
On each draw, a singleton is chosen with probability $\alpha b_s^{(1)}$ (cost
$+2$) and a path with probability at most $\alpha(n/s)$ (cost $+2R_s$); using
$N_s\alpha\le 1/2$ and multiplying the joint-hit sum by $\frac12$ yields
\begin{equation}\label{eq:round-credit}
 \E\!\left[A-\tfrac12 y(E[S_{N_s}])\right]
 \le b_s^{(1)}+\frac{R_s n}{s}-\frac{Z_s}{12} .
\end{equation}

\subsection{The gains at the two block lengths}\label{subsec:block-gains}

Write $E_{\rm out}$ for the arc set returned by \cref{alg:rounding}.

\paragraph{Long blocks ($s=\longsize$).}
Applying \cref{cor:fractional} directly with the cover-induced matching $y$
(which has matching deficit $n-y(E)=b_{\mathrm{long}}$) and using
$S_{N_s}\subseteq V_{\mathcal C}$ bounds the output deterministically by
\[
 |E_{\rm out}|\le\frac32 n+\initexcess+A+b_{\mathrm{long}}
                 -\frac12 y(E[S_{N_s}]),
\]
whose expectation via~\labelcref{eq:round-credit} saves $Z_{\mathrm{long}}/12$
against the singleton and return overheads.

\paragraph{Short blocks ($s=\shortsize$).}
For $s=\shortsize$, we modify $y$ after sampling to concentrate its weight
inside the sampled set $X=X_{N_s}$ and inside $V\setminus X$. First restrict
$y$ to $E[V\setminus U]$ by setting $w_{uv}=y_{uv}$ when $uv\in E[V\setminus U]$
and $w_{uv}=0$ for all other ordered pairs $(u,v)\in V\times V$. Because every
long path traversing $uv$ contains both $u$ and $v$, we have
$w_{uv}\le\rho_{\mathrm{long}}(u,v)$, so
$\|w\|_2^2=\sum_{u,v}w_{uv}^2\le Z_{\mathrm{long}}$; similarly, at each offset
two distinct vertices of a long path lie in at most one common short block,
giving $\rho_{\mathrm{short}}(u,v)\le\rho_{\mathrm{long}}(u,v)$ and
$Z_{\mathrm{short}}\le Z_{\mathrm{long}}$.

Write $I_v=\ind_{\{v\in X\}}$ (so $\E[I_v]=\beta$) and record at each tail $u$
and head $v$ the centered degree deviations into $X$:
\[
 D_u^+=\sum_{v\in V} w_{uv}(I_v-\beta),
 \qquad
 D_v^-=\sum_{u\in V} w_{uv}(I_u-\beta).
\]
For any $u,v\in V$ (including $u=v$, since $\rho_s(v,v)=1$), expressing
$I_uI_v$ via the probability $(1-2\alpha+\alpha\rho_s(u,v))^{N_s}$ of missing
both vertices across all $N_s$ draws gives
\[
 \Cov(I_u,I_v)=(1-2\alpha+\alpha\rho_s(u,v))^{N_s}
                   -(1-\alpha)^{2N_s}
 \le N_s\alpha\rho_s(u,v),
\]
using $(1-\alpha)^2\ge 1-2\alpha$ and convexity of $t\mapsto t^{N_s}$ on
$[0,1]$. Because $(\rho_s(v,v'))$ is symmetric with row sums
$\sum_{v'}\rho_s(v,v')\le s$ (the expected block size conditioned on hitting
$v$), applying $w_{uv}w_{uv'}\le\frac12(w_{uv}^2+w_{uv'}^2)$ yields
\[
 \E[(D_u^+)^2]
   =\sum_{v,v'}w_{uv}w_{uv'}\Cov(I_v,I_{v'})
   \le N_s\alpha\sum_{v,v'}\frac{w_{uv}^2+w_{uv'}^2}{2}\rho_s(v,v')
   \le N_s\alpha s\sum_{v\in V}w_{uv}^2,
\]
and symmetrically $\E[(D_v^-)^2]\le N_s\alpha s\sum_{u\in V}w_{uv}^2$. Summing
over $V$ and using $N_s\alpha\le 1/2$ gives
\begin{equation}\label{eq:round-variance}
 \sum_{v\in V}\E\!\left[(D_v^+)^2+(D_v^-)^2\right]
 \le 2N_s\alpha s\,\|w\|_2^2\le s\,\|w\|_2^2 .
\end{equation}

Now form a new fractional matching $y'$ on $E_r$: discard all entries of $w$
crossing $\delta(X)$, multiply entries inside $X\times X$ by $1/\beta$ and
entries inside $(V\setminus X)\times(V\setminus X)$ by $1/(1-\beta)$, and then
scale down any row sums above $1$ followed by any column sums above $1$. Before
trimming, the row sum at $u$ changes from its $w$-row sum by $D_u^+/\beta$ if
$u\in X$ and by $-D_u^+/(1-\beta)$ if $u\notin X$; since
$3/8\le\beta\le 1/2$, both $1/\beta$ and $1/(1-\beta)$ are at most $8/3<3$, so
row trimming removes at most $3|D_u^+|$ beyond any initial drop, and row
reductions can only decrease column excesses (which are bounded analogously by
$3|D_v^-|$). Because restricting $y$ to $E[V\setminus U]$ discarded at most
$2|U|$ weight, the deficit of $y'$ satisfies
\[
 n-y'(E)\le b_{\mathrm{long}}+2|U|+
                  3\sum_{v\in V}\bigl(|D_v^+|+|D_v^-|\bigr).
\]
Furthermore, $y'$ has no crossing arcs on $\delta(X)$ and at most $n-|X|$ total
weight on tails outside $X$, so $y'(E[X])\ge y'(E)-(n-|X|)=|X|-(n-y'(E))$.
Substituting this and $X\subseteq V_{\mathcal C}$ into \cref{cor:fractional},
and applying $\frac92|t|\le\frac{1}{64}+324t^2$ (from
$(18|t|-\frac18)^2\ge0$) to each of the $2n$ deviations $|D_v^+|,|D_v^-|$ in
$\frac32(n-y'(E))$, gives the deterministic bound
\[
 |E_{\rm out}|
 \le \frac32n+\initexcess+A+\frac32 b_{\mathrm{long}}+3|U|+\frac{n}{32}
     +324\sum_{v\in V}\bigl((D_v^+)^2+(D_v^-)^2\bigr)-\frac12|X|.
\]
To take expectations, write $\E[A]\le b_s^{(1)}+R_s n/s-Z_s/12+\frac12\E[y(E[S_{N_s}])]$
from~\labelcref{eq:round-credit}. Arcs of $y$ incident to $U$ carry weight at
most $2|U|$, while on $E[V\setminus U]$ our covariance bound gives
$\E[I_uI_v]\le\beta^2+N_s\alpha\rho_s(u,v)$ and thus
$\E[y(E[S_{N_s}])]\le 2|U|+\beta^2 n+N_s\alpha Z_s$. Combining this with
$\E|X|=\beta n$,~\labelcref{eq:round-variance},
$\frac12\beta(1-\beta)-\frac{1}{32}\ge\frac{15}{128}-\frac{1}{32}=\frac{11}{128}$
(for $\beta\in[3/8,1/2]$), and $\frac12 N_s\alpha-\frac{1}{12}\le\frac16$
yields
\begin{equation}\label{eq:round-transfer}
 \begin{aligned}
 \E[|E_{\rm out}|\mid s=\shortsize]
 &\le\frac32 n+\initexcess+\frac32 b_{\mathrm{long}}+b_s^{(1)}+\frac{R_s n}{s}\\
 &\quad+4|U|-\frac{11n}{128}+324s\,\|w\|_2^2+\frac{Z_s}{6} .
 \end{aligned}
\end{equation}

\paragraph{Balancing the two block lengths.}
Assign weight $\omega_{\mathrm{short}}=1/(1+8192\shortsize)$ to~\labelcref{eq:round-transfer}
and $\omega_{\mathrm{long}}=1-\omega_{\mathrm{short}}=8192\shortsize\,\omega_{\mathrm{short}}$
to the long-block bound. Using $\|w\|_2^2\le Z_{\mathrm{long}}$ and
$Z_{\mathrm{short}}\le Z_{\mathrm{long}}$, we have
\[
 324\shortsize\,\|w\|_2^2+\frac{Z_{\mathrm{short}}}{6}
 \le \left(324\shortsize+\frac16\right)Z_{\mathrm{long}}
 \le 8192\shortsize\cdot\frac{Z_{\mathrm{long}}}{12},
\]
so the long cover's joint-hit saving $-\omega_{\mathrm{long}}Z_{\mathrm{long}}/12$
absorbs the short cover's variance and joint-hit terms completely, leaving the
weighted average
\begin{equation}\label{eq:sample}
 \begin{aligned}
 \sum_{s\in\{\shortsize,\longsize\}}\omega_s\,\E[|E_{\rm out}|\mid s]
 &\le\frac32 n+\initexcess+b_{\mathrm{long}}
      +\sum_{s\in\{\shortsize,\longsize\}}\omega_s\left(b_s^{(1)}+\frac{R_s n}{s}\right)\\
 &\quad+\omega_{\mathrm{short}}\!\left(\frac{b_{\mathrm{long}}}{2}+4|U|-\frac{11n}{128}\right).
 \end{aligned}
\end{equation}

\subsection{A deterministic choice of blocks}\label{subsec:derandomize}

For a fixed size $s\in\{\shortsize,\longsize\}$ and any completed draw sequence,
the two deterministic upper bounds derived in \cref{subsec:block-gains} define
the estimator
\begin{equation}\label{eq:round-estimator}
 \begin{aligned}
 &\frac32 n+\initexcess+A\\[-2pt]
 &\quad+
 \begin{cases}
   b_{\mathrm{long}}-\tfrac12 y(E[S_{N_s}]),
        &s=\longsize,\\[3pt]
   \displaystyle \tfrac32 b_{\mathrm{long}}+3|U|+\tfrac n{32}
        +324\sum_{v\in V}\bigl((D_v^+)^2+(D_v^-)^2\bigr)-\tfrac12|X|,
        &s=\shortsize,
 \end{cases}
 \end{aligned}
\end{equation}
whose $\omega$-weighted expectation is at most the right-hand side
of~\labelcref{eq:sample}.

Moreover, after fixing any prefix of draws, the conditional expectation
of~\labelcref{eq:round-estimator} over the remaining independent draws can be
evaluated exactly in polynomial time. A vertex currently outside the prefix
remains unhit across $q$ independent draws with probability $(1-\alpha)^q$, and
two currently unhit vertices $u,v$ are both missed across $q$ draws with
probability $(1-2\alpha+\alpha\rho_s(u,v))^q$. By linearity of expectation,
these closed-form single- and pair-miss probabilities evaluate $\E|X|$,
$\E[y(E[S_{N_s}])]$, and every squared deviation $\E[(D_v^\pm)^2]$, as well as
every future expected charge in $A$: conditioned on choosing block $B$ at draw
$t+1$ (beyond the fixed prefix), an arc $uv\in E(B)$ is charged unless both $u$
and $v$ belong to $S_t$, and a vertex $v\in V(B)$ is discounted if $v\notin S_t$,
events determined solely by the intervening independent draws between the prefix
and step $t$.

We first evaluate the unconditional expectation of~\labelcref{eq:round-estimator}
at $s=\shortsize$ and $s=\longsize$, choose the size $s$ achieving the smaller
value (which is at most~\labelcref{eq:sample}), and then fix its $N_s$ draws
one by one, each time selecting a positive-weight block minimizing the
conditional expectation. Passing the resulting deterministic sequence to
\cref{alg:rounding} produces a spanning subgraph meeting the
bound~\labelcref{eq:sample}.

\subsection{The approximation ratio across all cases}\label{sec:parameters}

To see how the constants are chosen in~\labelcref{eq:sample}, note first that
the short cover contributes a saving of $-11n/128$ against a return cost of
$R_{\mathrm{short}}n/\shortsize$, both weighted by
$\omega_{\mathrm{short}}=\Theta(1/\shortsize)$. Fixing
$\xi_{\mathrm{short}}=1/64$ makes $R_{\mathrm{short}}=\Theta(\cycmin)$
by~\labelcref{eq:radius}, so taking $\shortsize$ to be a large constant
multiple of $\cycmin$ keeps $R_{\mathrm{short}}/\shortsize$ tiny and preserves a
weighted saving of order $n/\shortsize$. For the long cover,
choosing $\xi_{\mathrm{long}}=8/\cycmin^2$ gives
$R_{\mathrm{long}}=\Theta(\cycmin^3)$, so setting $\longsize=\cycmin^5$
bounds the return cost $R_{\mathrm{long}}n/\longsize$ by $O(n/\cycmin^2)$.
The matching deficit $b_{\mathrm{long}}$ is at most
$n/\longsize+b_{\mathrm{long}}^{(1)}$. By \cref{lem:path-cover}, the singleton
weight is at most $2\longsize(k+1)+2\xi_{\mathrm{long}}n$, where
$k=\OPT(G)-n$. Thus when $n\ge n_0=\longsize\cycmin^2$ and $k<n/n_0$, the
deficit is also $O(n/\cycmin^2)$.

Concretely, fix the input-independent parameters
\[
 \begin{aligned}
 \cycmin &= 2^{47}, &\qquad
 \shortsize &= 2^{20}\cycmin, &\qquad
 \xi_{\mathrm{short}} &= \frac{1}{64},\\
 \longsize &= \cycmin^5, &\qquad
 \xi_{\mathrm{long}} &= \frac{8}{\cycmin^2}, &\qquad
 n_0 &= \longsize\cycmin^2 = \cycmin^7,
 \end{aligned}
\]
define $R_{\mathrm{short}},R_{\mathrm{long}}$ via~\labelcref{eq:radius} and
$\joinradius=2\max(R_{\mathrm{short}},R_{\mathrm{long}})+\cycmin\longsize$, and
set $\hubcap=\lceil 5(\joinradius+2)\cycmin^2\rceil$, so each additive term
in~\labelcref{eq:merge} is at most $n/\cycmin^2$.

We now evaluate~\labelcref{eq:sample} assuming $n\ge n_0$, $k=\OPT(G)-n<n/n_0$,
and that neither long-cycle contraction nor either far-vertex test halted early.
At either size $s$, we have $b_s\ge n/s\ge n/\longsize\ge\cycmin^2=2^{94}\ge64$.
Combining~\labelcref{eq:initial-bounds,eq:merge} gives $|U|\le 101n/\cycmin$ and
$\initexcess\le 201n/\cycmin^2$. Since $(k+1)/n<1/n_0+1/n\le 2/n_0$ and
$\longsize/n_0=1/\cycmin^2$, \cref{lem:path-cover} (\labelcref{eq:singleton-bounds}) and
$b_s\le b_s^{(1)}+n/s$ give
\[
 \begin{aligned}
 \frac{b_{\mathrm{long}}^{(1)}}{n}
   &\le \frac{4\longsize}{n_0}+\frac{16}{\cycmin^2}
    = \frac{20}{\cycmin^2} < \frac{28}{\cycmin^2},\\
 \frac{b_{\mathrm{long}}}{n}
   &\le \frac{b_{\mathrm{long}}^{(1)}}{n}+\frac{1}{\longsize}
    < \frac{29}{\cycmin^2},\\
 \frac{b_{\mathrm{short}}^{(1)}}{n}
   &\le \frac{28}{\cycmin^2}+\frac{2\shortsize}{\longsize}+\frac{1}{32}
    \le \frac{1}{32}+\frac{1}{\cycmin}.
 \end{aligned}
\]
From~\labelcref{eq:radius}, $R_{\mathrm{long}}\le 65\cycmin^3$, so
$R_{\mathrm{long}}/\longsize\le 65/\cycmin^2$. Using $\omega_{\mathrm{long}}<1$,
the non-short terms
$\initexcess+b_{\mathrm{long}}+\omega_{\mathrm{long}}(b_{\mathrm{long}}^{(1)}+R_{\mathrm{long}}n/\longsize)$
in~\labelcref{eq:sample} sum to at most
$(201+29+28+65)n/\cycmin^2=323n/\cycmin^2$. Meanwhile,
$R_{\mathrm{short}}\le(2^{15}+14)\cycmin$, so the positive per-vertex terms
inside the $\omega_{\mathrm{short}}$-group satisfy
\[
 \frac{b_{\mathrm{long}}/2+4|U|+b_{\mathrm{short}}^{(1)}}{n}
 +\frac{R_{\mathrm{short}}}{\shortsize}
 \le \left(\frac{1}{32}+\frac{1}{32}\right)+\frac{405}{\cycmin}
     +\frac{29}{2\cycmin^2}+\frac{14}{2^{20}}
 <\frac{65}{1024}.
\]
Subtracting this from $11/128=88/1024$ leaves a net saving of more than
$23/1024$ per vertex before weighting. Because
$\omega_{\mathrm{short}}\ge 1/(8193\shortsize)$ and
$\cycmin^2/\shortsize=\cycmin/2^{20}=2^{27}$, the weighted short-cover saving
exceeds $\frac{23}{1024\cdot 8193}\cdot 2^{27}\,\frac{n}{\cycmin^2}>360\frac{n}{\cycmin^2}$,
giving a deterministic bound below $\frac32n-\frac{37n}{\cycmin^2}$.

\paragraph{Combining all cases.}
If $n<n_0$, we find an optimal SCSS by subset enumeration. For $n\ge n_0$, we
compute the rooted solution of \cref{lem:completion} ($\le\OPT(G)+n/2$) and
return the smaller of it and the subgraph produced by long-cycle contraction,
separator contraction at $s\in\{\shortsize,\longsize\}$, or shared-return
rounding.

If $k\ge n/n_0$, the rooted solution already uses at most
$\OPT(G)+n/2=\bigl(1+\frac{n}{2(n+k)}\bigr)\OPT(G)
\le\bigl(\frac32-\frac{1}{2(n_0+1)}\bigr)\OPT(G)$ arcs. If $k<n/n_0$, then
recall $n\le\OPT(G)<2n$ (\cref{sec:introduction}): reaching the contraction
threshold gives at most $\frac32n-n/\cycmin^2\le(\frac32-\frac{1}{2\cycmin^2})\OPT(G)$
arcs (\cref{lem:long}); triggering separator contraction at size $s$ gives at
most $\frac32\OPT(G)-\xi_s n/4\le(\frac32-\frac{1}{\cycmin^2})\OPT(G)$ arcs
(\cref{lem:metric}, as $\xi_s\ge 8/\cycmin^2$); and otherwise shared-return
rounding gives at most $\frac32n-37n/\cycmin^2\le(\frac32-\frac{37}{2\cycmin^2})\OPT(G)$
arcs. The tightest bound is the rooted threshold, yielding \cref{thm:main} with
\begin{equation}\label{eq:epsilon}
 \varepsilon_0=\frac{1}{2(n_0+1)}
 =\frac{1}{2(\cycmin^7+1)}=\frac{1}{2(2^{329}+1)}>2^{-331}.
\end{equation}

\paragraph{Running time.}
Discarding self-loops and parallel input arcs in $O(n^2+m)$ time leaves at most
$n(n-1)$ arcs. Computing $E_r$ takes $O(n^3)$ time, and bipartite matching,
minimum $s$--$t$ cuts, the ellipsoid separation of~\labelcref{eq:cut-lp}, and
minimum-price arborescences all run in polynomial time. There are at most $n-1$
cycle contractions, each building a polynomial-size cycle list; constructing
the tree decomposition, scanning $K$ bands and $L_{\rm depth}$ depth classes,
and solving pieces of bounded size~\labelcref{eq:piece-size} take polynomial
time for fixed parameters.

Finally,~\labelcref{eq:path-lp} lists $|\mathcal P|\le n^{\longsize}$ paths,
subdividing at $\shortsize$ offsets adds a constant factor, and derandomization
performs $N_s\le n$ sequential steps. At each step and candidate block, the
conditional expectation evaluates polynomially many rational powers of exponent
at most $n$ whose bases come from the LP solution, requiring polynomial bit
length and arithmetic time. Including exact enumeration for $n<n_0$, the total
running time is $n^{O(\longsize)}+O(m)$, completing the proof of
\cref{thm:main}.

\paragraph{AI Use Statement.}
The authors used OpenAI Astra in developing and revising the
manuscript, providing suggestions and critically checking the proposed arguments over successive exchanges. Gemini Argon assisted with some
parts of the work.

\begingroup
\small
\bibliographystyle{alpha}
\bibliography{scss-returns}
\endgroup

\appendix
\section{The gap in the charging argument for a reordered branch}\label[appendix]{app:bdk09gap}

Berman, DasGupta, and Karpinski~\cite{BermanDK09} grow and contract strongly
connected components along a directed depth-first search tree while maintaining
fractional credits on components to pay for closing arcs. In one configuration
of their $3/2$ analysis, a completed search branch must supply half a unit of
surplus credit to an enclosing triangle $A$. Case~3.3.1 in Appendix~A of their
full version (p.~12) establishes such a surplus whenever a closing branch
merges at least three components, each retaining at least $\frac12$ unit after
paying for its incoming tree arc, leaving at least $3\cdot\frac12-1=\frac12$
unit after buying the closing arc.

The gap arises in the final paragraph of Remark~1 (p.~13), which revisits
Case~3.3.5.1. There, an arc from the DFS subtree of one vertex $t_1$ of a
$2$-cycle (digon) $\{t_1,t_2\}$ enters a proper descendant of $t_2$ in the DFS
tree. To exploit this cross-arc, the authors retroactively assign the digon arc
$t_1t_2$ lowest search priority, altering the DFS traversal so that the return
path to $A$ visits $t_1$, $t_2$, and a component that lay strictly below $t_2$
in the original tree, and then invoke the three-component surplus of Case~3.3.1.
However, Case~3.3.1 requires $\frac12$ unit of \emph{remaining} balance on each
component present at the moment the branch closes, whereas Remark~1 only tracks
the initial credit a component inherits from its root: an intermediate contraction
along the modified branch may already have spent the credit of the component
below $t_2$. Closing the argument requires showing that the reordered traversal
pays for all arcs needed to merge the digon and its nested subtrees while still
leaving at least $\frac12$ unit for $A$.

\section{Sparse directed cuts and a branching approximation}\label[appendix]{app:sparse-cuts}

At an extreme point of a cut polyhedron, the active tight constraints must
uniquely determine every coordinate free to vary within its box bounds. After
uncrossing the tight cuts into a cross-free family, we count each underlying
vertex bipartition once and bound the additional rank contributed by reversing
cut orientations; Goemans~\cite{Goemans06} obtained the same $3n-4$ bound for
the asymmetric TSP relaxation using its flow-conservation equations.

\begin{proof}[Proof of \cref{lem:sparse}]
Fix $x\in P_h$ and put $D=\supp x$. Call a nonempty proper vertex subset
$S\subsetneq V$ a \emph{shore}, call $\{S,V\setminus S\}$ its
\emph{bipartition}, and say that $S$ is \emph{tight} if $x(\delta^-(S))=h$. For
each shore $S$, write $a_S\in\{0,1\}^D$ for the incidence vector of
$\delta^-(S)\cap D$. We bound the dimension of the linear subspace $W\subseteq\mathbb R^D$
spanned by $\{a_S : S\text{ is tight}\}$.

Following the directed-cut uncrossing argument of Melkonian and
Tardos~\cite{MelkonianTardos04}, choose a vector $c$ in the relative interior
of the convex cone generated by $\{a_S:S\text{ is tight}\}$ avoiding every
proper subspace of $W$ spanned by tight rows (there are finitely many). Among
all nonnegative representations $c=\sum_{S\text{ tight}}\lambda_S a_S$
($\lambda_S\ge0$), pick one maximizing $\sum_S\lambda_S|S|^2$; the feasible set
is compact since $h\sum_S\lambda_S=\sum_S\lambda_S(a_S\cdot x_D)=c\cdot x_D$.

Suppose two shores $S,T$ with $\lambda_S,\lambda_T>0$ \emph{cross}, meaning
that all four Venn regions $S\cap T$, $S\setminus T$, $T\setminus S$, and
$V\setminus(S\cup T)$ are nonempty. Then $S\cap T$ and $S\cup T$ are also
shores, so feasibility and submodularity give
$2h=x(\delta^-(S))+x(\delta^-(T))\ge x(\delta^-(S\cap T))+x(\delta^-(S\cup T))\ge 2h$,
forcing $S\cap T$ and $S\cup T$ to be tight as well. Moreover, the vector
$a_S+a_T-a_{S\cap T}-a_{S\cup T}$ is componentwise nonnegative and has inner
product zero with $x_D>0$, so it vanishes identically on $D$. Shifting
$\mu=\min\{\lambda_S,\lambda_T\}>0$ units of weight from $S$ and $T$ to
$S\cap T$ and $S\cup T$ preserves $\sum_U\lambda_U a_U=c$ while increasing
$\sum_U\lambda_U|U|^2$ by $2\mu|S\setminus T|\,|T\setminus S|>0$, a
contradiction. Thus the shores with $\lambda_S>0$ are pairwise cross-free, and
their rows span $W$ by our choice of $c$.

Fix a vertex $r\in V$ and represent each bipartition appearing with positive
weight by its unique shore $U\subseteq V\setminus\{r\}$ avoiding $r$. Because
$r\in V\setminus(U\cup U')$ for any two such shores $U,U'$, cross-freeness
implies that $U\cap U'=\varnothing$, $U\subseteq U'$, or $U'\subseteq U$; that
is, these distinct shores form a laminar family on $V\setminus\{r\}$, which has
size at most $2(n-1)-1=2n-3$. Keeping one row per bipartition contributes at
most $2n-3$ generators, and whenever both orientations $U$ and $V\setminus U$
are tight, their difference lies in the span of the $n$ singleton net-inflow
rows:
\[
 a_U-a_{V\setminus U}
   =\sum_{v\in U}\bigl(a_{\{v\}}-a_{V\setminus\{v\}}\bigr).
\]
Because $\sum_{v\in V}(a_{\{v\}}-a_{V\setminus\{v\}})=0$, those $n$ rows span a
subspace of dimension at most $n-1$, whence
$\dim W\le(2n-3)+(n-1)=3n-4$.

At an extreme point of $P_h$, the tight rows must span $\mathbb R^D$ (any
nonzero $d\in\mathbb R^D$ orthogonal to $W$ satisfies $x\pm\varepsilon d\in P_h$
for small $\varepsilon>0$), giving $|D|\le\dim W\le 3n-4$. Similarly, at an
extreme point of $P_h\cap[0,1]^E$, projecting the tight rows onto the
fractional coordinates $F=\{e\in E:0<x_e<1\}$ spans $\mathbb R^F$, giving
$|F|\le\dim W\le 3n-4$.
\end{proof}

\paragraph{Minimum-size $k$-arc-connected spanning subgraphs.}
Let $k\ge1$ be an integer, suppose each input arc may be selected at most once,
and let $x$ be an extreme-point optimum of $\min\{x(E):x\in P_k\cap[0,1]^E\}$
on a feasible instance. Write $X=x(E)$ and $F=\{e\in E:0<x_e<1\}$; the
singleton cuts $\delta^-(\{v\})$ give $X\ge kn$, and \cref{lem:sparse} gives
$|F|\le 3n-4$. Fix a root $r$. By Laekhanukit, Oveis Gharan, and
Singh~\cite[Corollary~4 and Remark~6]{LaekhanukitGS12}, one can compute in
polynomial time explicit distributions over packings of $k$ arc-disjoint
spanning in-arborescences and of $k$ arc-disjoint spanning out-arborescences
rooted at $r$, each using arc $e$ with marginal probability at most $x_e$.

Sample one packing of each orientation independently and let $H$ be the union
of all $2k$ arborescences. Every shore $S\subseteq V\setminus\{r\}$ receives at
least one arc from each of the $k$ out-arborescences, and every shore
$S\subsetneq V$ containing $r$ receives at least one arc from each of the $k$
in-arborescences, so $H$ is $k$-arc-connected. By independence (and integrality
outside $F$), $\Prb(e\in H)\le 1-(1-x_e)^2=2x_e-x_e^2$, whence
\[
 \E|H|\le X+\sum_{e\in F}x_e(1-x_e).
\]
For $t\in[0,1]$, both $t(1-t)\le 1/4$ and $t(1-t)=\frac19+\frac{t}{3}-(t-\frac13)^2\le\frac{t}{3}+\frac19$
hold. The first inequality bounds the sum over $F$ by
$|F|/4\le 3n/4\le 3X/(4k)$, and the second bounds it by
$X/3+|F|/9\le X/3+n/3\le 2X/3$. Enumerating all pairs of packings in the
polynomial-size supports deterministically achieves at most
$\min\{\frac53,\,1+\frac{3}{4k}\}X$, and hence a
$\min\{\frac53,\,1+\frac{3}{4k}\}$-approximation.

\end{document}